\documentclass[usenatbib]{mnras}

\usepackage{newtxtext,newtxmath}

\usepackage[T1]{fontenc}
\usepackage{times,ae,aecompl}

\usepackage{graphicx}	
\usepackage{amsmath}	
\usepackage{amssymb}	
\usepackage{wasysym}
\usepackage{xcolor}
\usepackage{multirow}
\usepackage{pdfcolfoot}

\usepackage{etoolbox}
\makeatletter
\patchcmd\@combinedblfloats{\box\@outputbox}{\unvbox\@outputbox}{}{\errmessage{\noexpand patch failed}}
\makeatother

\newcommand{\bq}{\begin{eqnarray}}
\newcommand{\eq}{\end{eqnarray}}
\newcommand{\be}{\bq}
\newcommand{\ee}{\eq}
\newcommand{\hMpc}{$h \, {\rm Mpc}^{-1}$}
\newcommand{\Mpch}{{\rm Mpc}\,h^{-1}}
\newcommand{\fRbg}{\bar{f}_{R0}}
\newcommand{\diff}{{\rm d}}

\newcommand{\smallsub}{\rm \scriptscriptstyle}

\newcommand{\ijmpd}{Int. J. Mod. Phys. D}

\definecolor{darkgreen}{cmyk}{0.85,0.2,1.00,0.05}

\title[Matter power spectrum reaction to dark energy and modified gravity]{On the road to percent accuracy: nonlinear reaction of the matter power spectrum to dark energy and modified gravity}

\author[Cataneo et al.]{M. Cataneo$^{1}$\thanks{E-mail: matteo@roe.ac.uk}, 
L. Lombriser$^{1,2}$,
C. Heymans$^{1}$,
A. J. Mead$^{3,4}$,
A. Barreira$^{5}$,\newauthor
S. Bose$^{6}$,
B. Li$^{7}$
\\
$^{1}$Institute for Astronomy, University of Edinburgh, Royal Observatory, Blackford Hill, Edinburgh, EH9 3HJ, U.K.\\
$^{2}$D\'epartement de Physique Th\'eorique, Universit\'e de Gen\`eve, 24 quai Ernest Ansermet, 1211 Gen\`eve 4, Switzerland\\
$^{3}$Department of Physics and Astronomy, University of British Columbia, Vancouver, BC V6T 1Z1, Canada\\
$^{4}$ICC, Instituto de Ciencias del Cosmos, Universitat de Barcelona, IEEC-UB, Mart\'i i Franqu\`es 1, E-08028, Barcelona, Spain\\
$^{5}$Max-Planck-Institut f\"{u}r Astrophysik, Karl-Schwarzschild-Str. 1, 85741 Garching, Germany\\
$^{6}$Institute for Theory and Computation, Harvard Smithsonian Center for Astrophysics, 60 Garden Street, Cambridge, MA02138, USA\\
$^{7}$Institute for Computational Cosmology, Department of Physics, Durham University, Durham DH1 3LE, U.K.\\
}

\date{Accepted XXX. Received YYY; in original form ZZZ}

\pubyear{2018}

\begin{document}
\label{firstpage}
\pagerange{\pageref{firstpage}--\pageref{lastpage}}
\maketitle

\begin{abstract}
We present a general method to compute the nonlinear matter power spectrum for dark energy and modified gravity scenarios with percent-level accuracy. By adopting the halo model and nonlinear perturbation theory, we predict the reaction of a $\Lambda$CDM matter power spectrum to the physics of an extended cosmological parameter space. By comparing our predictions to $N$-body simulations we demonstrate that with {\it no-free parameters} we can recover the nonlinear matter power spectrum for a wide range of different $w_0$-$w_a$ dark energy models to better than 1\% accuracy out to $k \approx 1$~\hMpc. We obtain a similar performance for both DGP and $f(R)$ gravity, with the nonlinear matter power spectrum predicted to better than 3\% accuracy over the same range of scales. When including direct measurements of the halo mass function from the simulations, this accuracy improves to 1\%. With a single suite of standard $\Lambda$CDM $N$-body simulations, our methodology provides a direct route to constrain a wide range of non-standard extensions to the concordance cosmology in the high signal-to-noise nonlinear regime.
\end{abstract}




\section{Introduction}

%


General Relativity (GR) has been put under intense scrutiny in the Solar System, where it has successfully passed all tests~\citep{Will:2014kxa}. Its application to cosmology, however, involves vastly different length scales and is comparable in orders of magnitude to an extrapolation from an atomic nucleus to the scale of human experience.
It is therefore important to perform independent tests of our Theory of Gravity in the cosmological regime.
Further motivation for a thorough inspection of cosmological gravity can be drawn from the necessity of a large dark sector in the energy budget of our Universe to explain large-scale observations with GR~\citep{Riess:1998cb,Perlmutter:1998np,Aghanim:2018eyx,Hildebrandt:2016iqg,Abbott:2017wau}.
In particular the late-time accelerated expansion of the cosmos has traditionally been an important driver for the development of alternative theories of gravity, a concept that has however become strongly challenged with the confirmation of the luminal speed of gravity~\citep{Lombriser:2016a,Abbott:2017b,Lombriser:2017a,Creminelli:2017,Ezquiaga:2017,Baker:2017,Sakstein:2017xjx,Battye:2018ssx,deRham:2018red,Creminelli:2018xsv}.
Nevertheless, cosmic acceleration could be the result of a dark energy field permeating the Universe that may well be coupled to matter with an observable impact on cosmological scales.
Importantly, should that coupling be universal, i.e. affecting baryons and dark matter equally, the corresponding models must then rely on the employment of screening mechanisms to comply with the stringent Solar-System bounds~\citep{Vainshtein:1972sx,Khoury:2004,Babichev:2009ee,Hinterbichler:2010es}.
Signatures of screening are naturally to be expected in the nonlinear cosmological small-scale structure, where modified gravity transitions to GR, and for some models are even exclusively confined to these scales~\citep{Wang:2012kj,Heymans2018}.
The increasing wealth of high-quality data at these scales~\citep{Laureijs:2011,LSST:2012,Hildebrandt:2016iqg,Abbott:2017wau} renders cosmological tests of gravity a very timely enterprise.
At the same time, cosmological structure formation proves notoriously difficult to model to sufficient accuracy in this regime, where high signal-to-noise measurements have the potential to distinguish a few percent deviation from GR~\citep{Heymans2018}.

For any given theory of gravity or dark energy model, our current best predictions for the statistical properties of the resulting matter distribution come from large-volume high-resolution \mbox{$N$-body} simulations~\citep{Oyaizu:2008,Schmidt:2009a,Zhao:2011,Li:2011vk,Brax:2012b,Baldi:2012,Puchwein:2013,Wyman:2013,Barreira:2013,Li:2013tda,Llinares:2014,Winther:2015b}. Running these, however, can take up to thousands of node-hours on dedicated cluster facilities, and although methods to partially alleviate this drawback exist~\citep[see, e.g.,][]{Barreira:2015,Mead:2015b,Valogiannis:2017,Winther:2017,Bose:2017,Llinares:2017} exploring vast swathes of the theory space remains currently unfeasible. Alternatively, analytical and semi-analytical methods can be used to swiftly predict specific large-scale structure observables, such as the matter power spectrum~\citep{Koyama:2009,Schmidt:2009c,Schmidt:2010,Li:2011,Fedeli:2012,Brax:2013,Lombriser:2014,Barreira:2014a,Barreira:2014b,Zhao:2014,Achitouv:2016,Mead:2016,Aviles:2017,Cusin:2018,Bose:2018,Hu:2018}, with the important caveat that they have limited accuracy in the nonlinear regime of structure formation, and often involve some level of fitting to the same quantity measured in simulations. These approaches are therefore inadequate for future applications to high-quality data from Stage IV surveys~\citep{Laureijs:2011,LSST:2012,Levi:2013,Koopmans:2015}, where percent level accuracy over a wide range of scales will be necessary to obtain tight and unbiased constraints on departures from GR~\citep{Alonso:2017,Casas:2017,Reischke:2018,Mancini:2018,Taylor:2018b} and the nature of dark energy~\citep{Albrecht:2006}. Matter power spectrum emulators can provide a solution to this problem for particular modified gravity or dark energy models~\citep{Heitmann:2014,Lawrence:2017,Knabenhans:2018,Winther:2019}, but still rely on the availability of large quantities of computational resources to the determine the properties of the matter power spectrum at the location of the emulator nodes. The absence of a clear attractive alternative to the $\Lambda$CDM paradigm calls for a more general framework, one easily adaptable to non-standard cosmologies beyond the handful of well studied cases. 

Here we take an important step in this direction by extending the method proposed in~\cite{Mead:2017}, where the halo model is used to compute matter power spectrum ratios with respect to a convenient baseline cosmology. \cite{Mead:2017} showed that by determining these ratios, rather than the absolute value of the matter power spectrum, the shortcomings of the halo model are mitigated. The initial conditions of the baseline cosmology are designed so that under GR+$\Lambda$CDM evolution the linear clustering of matter at some given redshift exactly reproduces that of the target cosmology of interest, whose evolution is instead governed by non-standard laws of gravity and/or background expansion. Assuming one can generate an accurate nonlinear matter power spectrum for the reference cosmology (e.g with a suitable emulator), recovery of the target power spectrum then hinges on the computation of a `correction' factor that incorporates the nonlinear effects of fifth forces, screening mechanisms and deviations from the cosmological constant. We use the halo model and nonlinear perturbation theory to obtain such corrections, and refer to this quantity as the \emph{reaction}.

The paper is organised as follows. In Sec.~\ref{sec:DE_MG} we briefly describe popular modified gravity and dark energy models used here as testbeds for our methodology. Sec.~\ref{sec:method} reviews the halo model formalism and introduces the matter power spectrum reactions. The cosmological simulations used to validate our approach are described in Sec.~\ref{sec:sims}, and Sec.~\ref{sec:results} presents the capability of the halo model reactions to predict the nonlinear matter power spectrum. We summarise our conclusions in Sec.~\ref{sec:conclusions}. 
We provide details of the spherical collapse and perturbation theory calculations employed in this work in App.~\ref{sec:SC} and App.~\ref{sec:PT}, respectively. Additional tests to gauge the importance of the halo mass function and halo concentration in our predictions are presented in App.~\ref{sec:tests}.

\section{Dark energy and modified gravity theory}
\label{sec:DE_MG}

The most general four dimensional scalar-tensor theory with second-order equations of motion is described by the action \citep{Horndeski:1974,Deffayet:2009,Kobayashi:2011}
\begin{eqnarray}\label{eq:L_horndeski}
S=\int \diff^4x \sqrt{-g}\left\{\sum_{i=2}^5{\cal L}_i[\phi,g_{\mu\nu}]+{\cal L}_{\rm m}[{\psi},g_{\mu\nu}]\right\} \, ,
\end{eqnarray}
where $g$ is the determinant of the metric $g_{\mu\nu}$ minimally coupled to a generic matter field $\psi$ (Jordan frame), ${\cal L}_{\rm m}$ is the matter Lagrangian, $\phi$ is the scalar degree of freedom, and the terms entering the Einstein-Hilbert Lagrangian are
\begin{eqnarray}
{\cal L}_2&=& K(\phi,X) \, ,  \nonumber \\
{\cal L}_3&=&  -G_3(\phi,X) \Box\phi \, , \nonumber \\
{\cal L}_4&=&   G_4(\phi,X)R+G_{4X}(\phi,X)\left\{(\Box \phi)^2-\nabla_\mu\nabla_\nu\phi \nabla^\mu\nabla^\nu\phi\right\} \, , \nonumber \\
{\cal L}_5&=& G_5(\phi,X)G_{\mu\nu}\nabla^\mu\nabla^\nu\phi
-\frac{1}{6}G_{5X}(\phi,X)\big\{ (\nabla\phi)^3 \nonumber \\ & & 
-3\nabla^\mu\nabla^\nu\phi\nabla_\mu\nabla_\nu\phi\Box\phi  
+2\nabla^\nu\nabla_\mu\phi \nabla^\alpha\nabla_\nu\phi\nabla^\mu\nabla_\alpha \phi
\big\}   \,.
\end{eqnarray}
Here, $K$ and $G_i$ are arbitrary functions of $\phi$ and $X\equiv-\nabla^\nu\phi\nabla_\nu\phi/2$, and the subscripts $X$ and $\phi$ denote derivatives. 

The nearly simultaneous detection of gravitational waves and electromagnetic signals emitted from two colliding neutron stars \citep{Abbott:2017a} imposes tight constraints on the present-day speed of gravitational waves $c_T$, i.e. $|c_T/c-1| \lesssim 10^{-15}$, where $c$ is the speed of light \citep{Abbott:2017b}. Restricting ourselves to theories of gravity with non-evolving $c_T$, and requiring this not to be achieved by extreme fine tuning of the $G_4$ and $G_5$ functions \citep{Lombriser:2016a,Lombriser:2017a,Creminelli:2017,Ezquiaga:2017,Baker:2017,Sakstein:2017xjx,Battye:2018ssx,deRham:2018red}, implies that the remaining Horndeski Lagrangian takes the form \citep{McManus:2016kxu}
\be\label{eq:reduced_horndeski}
\mathcal{L}_{\rm H} = K(\phi,X) + G_4(\phi)R - G_3(\phi,X) \Box\phi \, . 
\ee
In this paper we focus on well-studied models of modified gravity and dark energy, each exploring the effects introduced by the individual terms in Eq.~\eqref{eq:reduced_horndeski}.
Quintessence and k-essence dark energy models are described by a contribution of $K$ only (Sec.~\ref{sec:DE}). $G_4$ introduces a coupling of this field to the metric that modifies gravity.
The class of models described by this term encompasses the chameleon~\citep{Khoury:2004}, symmetron~\citep{Hinterbichler:2010es}, and k-mouflage~\citep{Babichev:2009ee} screening mechanisms, and we will study a particular example of this action with chameleon screening in Sec.~\ref{sec:fR} when considering a realisation in $f(R)$ gravity.
The $G_3$ term appears, for instance, in the four-dimensional effective scalar-tensor theory of DGP braneworld gravity (Sec.~\ref{sec:DGP}) and gives rise to the Vainshtein screening mechanism~\citep{Vainshtein:1972sx}.
Either $G_3$ or noncanonical kinetic contributions in $K$ produce a nonluminal sound speed of the scalar field fluctuations that can yield observable scale-dependent effects beyond the sound horizon.
The mass scale associated with a scalar field potential in $K$ can furthermore introduce a scale-dependent growth of structure below the sound horizon.
With $c_T=1$, a genuine self-acceleration of the cosmological background that is directly attributed to modified gravity must arise from $G_4$~\citep{Lombriser:2016a}, which is however in tension with observations~\citep{Lombriser:2017a}.
A self-acceleration from $K$ or $G_3$ that dispenses with the need of a cosmological constant is, in contrast, still observationally feasible.

Throughout, we assume a flat Friedmann-Robertson-Walker (FRW) background, and the perturbed metric in the Newtonian gauge reads\footnote{Here and throughout we work in natural units, and set $c=1$.}
\be\label{eq:FRW}
\diff s^2 =g_{\mu\nu}\diff x^\mu \diff x^\nu = -(1+2\Psi)\diff t^2 + a^2(1+2\Phi)\diff {\bf x}^2 \, ,
\ee
where $\Psi$ and $\Phi$ denote the two gravitational potentials, and $a$ is the scale factor. The evolution of non-relativistic matter perturbations is determined by $\Psi$, whereas photons follow the null geodesics defined by the lensing potential $\Phi_- = (\Psi - \Phi)/2$~\citep[see, e.g.,][]{Carroll:2004}. For all models considered here $\Phi_- = \Psi_N$, where $\Psi_N$ is the standard Newtonian potential.

\subsection{$f(R)$ gravity} \label{sec:fR}

In $f(R)$ gravity the Einstein-Hilbert action is modified to contain an additional non-linear function of the Ricci scalar $R$, that is
\be\label{eq:fR_action}
S_{\rm EH}=\int \diff^4 x \sqrt{-g} \frac{1}{16\pi G}\left[R + f(R) \right] \, . 
\ee
The $f(R)$ Lagrangian is a particular case of Eq.~\eqref{eq:reduced_horndeski} with the Horndeski functions~\citep[see, e.g.,][]{deFelice:2011}
\be
K &=& -\frac{1}{16\pi G}[R f_R-f] \, ,\\
G_3 &=& 0 \, ,\\
G_4 &=& \frac{1}{16\pi G}(1+f_R) \, ,
\ee
where we defined the scalaron field $f_R \equiv \diff f/\diff R$, and used $\phi = (1 + f_R)/\sqrt{8\pi G}$. In the quasi-static regime\footnote{See, e.g., \cite{Noller:2014}, \cite{Bose:2015} and \cite{Lagos:2018} for a detailed discussion on the validity of the quasi-static approximation in modified gravity.}, structure formation is governed by the following coupled equations~\citep[e.g.,][]{Oyaizu:2008}
\bq
\label{eq:poisson-fr}\nabla^2\Psi &=& \frac{16\pi G}{3}\delta\rho_{\rm m} - \frac{1}{6}\delta R(f_R) \, , \\
\label{eq:eom-fr}\nabla^2 \delta f_R &=& \frac{1}{3}\left[\delta R(f_R) - 8\pi G\delta\rho_{\rm m}\right] \, ,
\eq
where $\delta\rho_{\rm m} = \rho_{\rm m} - \bar{\rho}_{\rm m}$, $\delta R = R - \bar{R}$ and $\delta f_R = f_R - \bar{f}_R$ are the matter density, curvature and scalaron perturbations with respect to their background averaged values. Eqs.~\eqref{eq:poisson-fr} and~\eqref{eq:eom-fr} can be combined to give 
\be\label{eq:gradientmod-fr}
\nabla \Psi = \nabla \Psi_{\rm N} - \frac{1}{2} \nabla \delta f_R \, ,
\ee
which explicitly shows that the scalar field fluctuations source an additional fifth force.

Since GR accurately describes gravity in our Solar System, viable modifications must also be compatible with local constraints. In $f(R)$ gravity this is achieved by means of the chameleon screening mechanism~\citep{Khoury:2004}, which suppresses departures from standard gravity for large enough potential wells $\Psi_{\rm N}$. In practice, structures are screened if the {\it thin shell} condition,
\be\label{eq:thin_shell}
|\delta f_R| \ll \frac{2}{3} |\Psi_{\rm N}| \, ,
\ee
is satisfied. Stable theories require $f_R < 0$ \citep{Hu:2007}, thus the chameleon screening activates throughout an isolated object if $|\delta f_R| \leq |\bar f_R| \ll |\Psi_{\rm N}|$. Assuming that the Milky Way is placed in the cosmological background, and knowing that $|\Psi_{\rm MW}| \sim 10^{-6}$, this in turn imposes $|\bar f_{R0}| < 10^{-6}$ for the present value of the background scalaron field. 

Hereafter, we adopt the following $f(R)$ functional form \citep{Hu:2007}
\be\label{eq:hs}
f(R) = -2\Lambda - \bar f_{R0} \frac{\bar R_0^2}{R} \, ,
\ee
where $\Lambda$ is an effective cosmological constant driving the background cosmic acceleration, and $\bar R_0$ corresponds to the background Ricci scalar today. We will work with values $|\bar f_{R0}| = 10^{-5}$ (F5) and $|\bar f_{R0}| = 10^{-6}$ (F6), for which cosmological structures are, respectively, partially unscreened or largely screened throughout the cosmic history. Note that deviations from the $\Lambda$CDM expansion history are of order $\bar f_{R0}$~\citep{Hu:2007}. Hence, for the $f(R)$ models considered here the background evolution is in effect equivalent to that of the concordance cosmology, with the Hubble parameter given by
\be\label{eq:Friedmann}
H^2 = \frac{8\pi G}{3}(\bar\rho_{\rm m} + \bar\rho_\Lambda) \, ,
\ee
where $\bar\rho_\Lambda$ is the energy density of the cosmological constant. The large-scale structure data currently available allows amplitudes $|\bar f_{R0}| \lesssim 10^{-5}$~\citep[][]{Terukina:2013eqa,Lombriser:2014dua,Cataneo:2015,Liu:2016,Alam:2016}, thus placing the F5 model on the edge of the region of parameter space still relevant for cosmological applications\footnote{See, however, the recent work by \cite{He:2018} where it was showed that deviations as small as $|\bar f_{R0}| =10^{-6}$ could already be in strong tension with redshift space distortions data. At the present time, the tightest constraints on $f(R)$ gravity come from the analysis of kinematic data for the gaseous and stellar components in nearby galaxies, which only allows $|\bar f_{R0}| \lesssim 10^{-8}$~\citep{Desmond:2018euk}.} if other effects degenerate with the enhanced growth of structure are ignored. Accounting for massive neutrinos~\citep{Baldi:2014} and baryonic feedback~\citep{Puchwein:2013,Hammami:2015,Arnold:2018} will loosen the existing constraints~\citep[see, e.g.,][]{Hagstotz:2018,Giocoli:2018}. In addition, alternative functional forms to Eq.~\eqref{eq:hs} can lead to different upper bounds on $|\bar f_{R0}|$~\citep[see, e.g.,][]{Cataneo:2015}.

\subsection{DGP} \label{sec:DGP}

In the DGP braneworld model the matter fields live on a four-dimensional brane embedded in a five-dimensional Minkowski space~\citep{Dvali:2000}. In this model the dimensionality of the gravitational interaction is controlled by the {\it crossover scale} parameter $r_{\rm c}$, such that on scales smaller than $r_{\rm c}$ DGP becomes a four-dimensional scalar-tensor theory described by an effective Lagrangian with terms~\citep{Nicolis:2004,Park:2010}
\be
K &\sim& r_{\rm c}^2 X^2 \, ,\label{eq:kinetic_DGP}\\
G_3 &\sim& r_{\rm c}^2 X \, ,\label{eq:G3_DGP}\\
G_4 &=& \frac{1}{16\pi G}e^{-\sqrt{\frac{16\pi G}{3}}\varphi} \, , \label{eq:G4_DGP}
\ee
where the brane-bending mode $\varphi$ represents the scalar field. Hereafter we will be working with the {\it normal branch} DGP model (nDGP), which despite being a stable solution of the theory is also incompatible with the observed late-time cosmic acceleration. To obviate this problem the Lagrangian given by Eqs.~\eqref{eq:kinetic_DGP}-\eqref{eq:G4_DGP} is extended to include a smooth, quintessence-type dark energy with a potential conveniently designed to match the expansion history of a flat $\Lambda$CDM cosmology~\citep{Schmidt:2009a}\footnote{This is an assumption made to ease comparisons to $\Lambda$CDM simulations, and is not a strict observational requirement~\citep[cf.][]{Lombriser:2009xg}.}. Therefore, the Friedmann equation Eq.~\eqref{eq:Friedmann} applies here as well.

The scalar field $\varphi$ couples to non-relativistic matter by sourcing the dynamical potential $\Psi$, which in turn produces a gravitational force given by 
\be\label{eq:gradientmod-DGP}
\nabla\Psi = \nabla\Psi_{\rm N} + \frac{1}{2} \nabla\varphi \, ,
\ee
where the second term on the right hand side is the attractive fifth force contribution. On length scales $\lambda \ll H^{-1}, r_{\rm c}$, and in the quasi-static regime~\citep{Schmidt:2009b,Brito:2014,Winther:2015}, the evolution of the brane-bending mode is described by~\citep{Koyama:2007}
\be\label{eq:phiQS}
\nabla^2 \varphi + \frac{r_{\rm c}^2}{3\beta} \left[ (\nabla^2\varphi)^2
- (\nabla_i\nabla_j\varphi)(\nabla^i\nabla^j\varphi) \right] = \frac{8\pi G}{3\beta} \delta\rho_{\rm m} \, ,
\ee
with the function $\beta(a)$ defined as
\be\label{eq:beta}
\beta(a) \equiv 1 + 2 H\, r_{\rm c} \left [ 1 + \frac{\dot H}{3 H^2} \right ] \, ,
\ee
where overdots denote derivatives with respect to cosmic time. The derivative self-interactions in Eq.~\eqref{eq:phiQS} suppress the field in high-density regions, where the matter density field is nonlinear, effectively restoring GR. This is the so-called Vainshtein screening. To explicitly illustrate how this mechanism works we shall consider a spherically symmetric overdensity with mass
\be
\delta M(r) = 4\pi \int_0^r \diff r^\prime r^{\prime 2} \delta\rho_{\rm m}(r^\prime) \, .
\ee
Then, for this system the gradient of $\varphi$ reads~\citep{Koyama:2007,Schmidt:2010}
\be
\frac{\diff\varphi}{\diff r} = \frac{4}{3\beta}\left(\frac{r}{r_{\rm V}}\right)^3\left[ \sqrt{1 + \left(\frac{r_{\rm V}}{r}\right)^3} - 1 \right]\frac{G \delta M(r)}{r^2} \, ,
\ee
where we defined the Vainshtein radius
\bq\label{eq:rv}
r_{\rm V}(r) \equiv \left[\frac{16r_{\rm c}^2G \delta M(r)}{9\beta^2}\right]^{1/3} \, .
\eq
The scale introduced in Eq.~\eqref{eq:rv} sets the distance from the centre of the spherical mass distribution above which fifth force effects are observable. For instance, for a top-hat overdensity of radius $R_{\smallsub TH}$ one has the two limiting cases
$$
\left\{
\begin{array}{ll}
\frac{\diff \varphi}{\diff r} = \frac{2}{3\beta}\frac{\diff\Psi_{\rm N}}{\diff r} & \quad r \gg r_{\rm V}> R_{\smallsub TH} \, ,\label{eq:scr_th} \\
\frac{\diff \varphi}{\diff r} \approx 0 & \quad R_{\smallsub TH} < r \ll r_{\rm V} \, . \label{eq:unscr_th}
\end{array}
\right.
$$

In the following we will consider the \emph{medium} and \emph{weak} nDGP variants used in~\cite{Barreira:2016}, with crossover scales $r_{\rm c}H_0 = 0.5$ (nDGPm) and $r_{\rm c}H_0 = 2$ (nDGPw) in units of the present-day Hubble horizon $H_0^{-1}$. Note that, at present, of these two cases nDGPw is the only one compatible with growth rate data~\citep{Barreira:2016}. Hence, similarly to F5, the use of nDGPm will serve as a testbed for our methodology in conditions of relatively strong departures from GR.

\subsection{Dark energy}\label{sec:DE}

The simplest models described by the Lagrangian in Eq.~\eqref{eq:reduced_horndeski} are those in which the scalar field is minimally coupled to gravity, that is
\be
K &=& K(\phi,X) \, ,\\
G_3 &=& 0 \, ,\\
G_4 &=& \frac{1}{16\pi G} \, .
\ee
In this scenario, the field $\phi$ is associated with a fluid called dark energy (DE) with energy density and pressure~\citep[see, e.g.,][]{Amendola:2010}
\be
\rho_{\rm \scriptscriptstyle DE} &=& 2 X K_X - K \, ,\\
P_{\rm \scriptscriptstyle DE} &=& K \, ,
\ee
respectively. Its background evolution is controlled by the equation of state parameter $w \equiv \bar P_{\rm \scriptscriptstyle DE}/\bar \rho_{\rm \scriptscriptstyle DE}$, and the solution to the continuity equation
\be
\frac{\diff  \bar \rho_{\rm \scriptscriptstyle DE}}{\diff t} + 3 H \bar \rho_{\rm \scriptscriptstyle DE}(1+w) = 0
\ee
is given by
\be\label{eq:DEevo}
\bar\rho_{\smallsub DE}(a) = \bar\rho_{{\smallsub DE},0} \exp \left[ 3\int_a^1 \frac{\diff a^\prime}{a^\prime} (1 + w) \right] \, ,
\ee
where $\bar\rho_{{\smallsub DE},0}$ is the present-day dark energy density.

Popular models of dark energy, such as quintessence~\citep{Wetterich:1988,Ratra:1988}, k-essence~\citep{Armendariz:2000} and clustering quintessence~\citep{Creminelli:2009}, belong to this subclass of theories. In this paper we restrict our discussion to a quintessence-like dark fluid with rest-frame sound speed $c_{\rm s}^2=1$ and equation of state~\citep{Chevallier:2001,Linder:2003}
\be\label{eq:CPL}
w(a) = w_0 + (1-a)w_a \, ,
\ee 
where $\{w_0,w_a\}$ are free phenomenological parameters. The relativistic sound speed washes out the dark energy perturbations on sub-horizon scales, resulting in modifications to the growth of structure tied solely to the different expansion history compared to $\Lambda$CDM. Our methodology can, in principle, also be applied to forms of dark energy clustering on small scales. 

Table~\ref{tab:de} summarises the dark energy models selected for this work, which have been chosen to roughly enclose the 2$\sigma$ region of parameter space allowed by the {\it Planck}~2015 temperature and polarization data in combination with baryon acoustic oscillations, supernova Ia and $H_0$ measurements~\citep{Planck:2016a}. 

\begin{table}
\begin{center}
\caption{Equation of state parameters defining the dark energy models used in this work.}
\begin{tabular}{c c c c}
\hline\hline
Model & $w_0$ & $w_a$ \\
\hline
DE1 & $-0.7$ & $0$  \\
DE2 & $-1.3$ & $0$  \\
DE3 & $-1$ & $0.5$  \\
DE4 & $-1$ & $-0.5$  \\
DE5 & $-0.7$ & $-1.5$  \\
DE6 & $-1.3$ & $0.5$  \\
\hline\hline
\end{tabular}
\label{tab:de}
\end{center}
\end{table}

\section{Matter power spectrum reaction}\label{sec:method}

In Sec.~\ref{sec:sphericalcollapse} and Sec.~\ref{sec:hm} we briefly review the spherical collapse model and the halo model formalism, which we use to predict the nonlinear matter power spectrum for the range of cosmological models listed in Sec.~\ref{sec:DE_MG}. The halo model assumes that all matter in the Universe is localized in virialized structures, called halos. In this approach, the spatial distribution of these objects and their density profiles determine the statistics of the matter density field on all scales. It is typically assumed that each mass element belongs to one halo only, i.e. halos are spatially exclusive. Below we introduce the ingredients entering the halo model prescription, and refer the interested reader to the~\cite{Cooray:2002} review on the topic for more details. In Sec.~\ref{sec:reactions} we then detail our new approach to reach percent level accuracy on these power spectra over a range of scales where the halo model alone is known to fail.

\subsection{Spherical collapse model} \label{sec:sphericalcollapse}

The Press-Schechter formalism~\citep{Press:1974} approximates halo formation following the evolution of a spherical top-hat overdensity of radius $R_{\smallsub TH}$ and mass $M=4\pi R_{\smallsub TH}^3 \bar\rho_{\rm m}(1+\delta)/3$ in an otherwise homogenous background. Mass conservation and the Euler equations imply~\citep[see, e.g.,][]{Schmidt:2009c}
\be\label{eq:SC1}
\frac{\ddot R_{\smallsub TH}}{R_{\smallsub TH}} = -\frac{4\pi G}{3}[\bar\rho_{\rm m} + (1+3w)\bar\rho_{\rm eff}] - \frac{1}{3}\nabla^2\Psi \, .
\ee
Here, $\bar\rho_{\rm eff}$ and $w$ are, respectively, the background energy density and equation of state of an effective dark energy component causing the late-time cosmic acceleration. Hence, in $f(R)$ gravity and nDGP, $\bar\rho_{\rm eff} = \bar\rho_\Lambda$ and $w=-1$. For the smooth dark energy models in Sec.~\ref{sec:DE} we have $\bar\rho_{\rm eff} = \bar\rho_{\smallsub DE}$ and $w$ given by Eq.~\eqref{eq:CPL}. Modifications of gravity enter through the potential term in Eq.~\eqref{eq:SC1}, which we parametrize as
\be\label{eq:param_poisson}
\nabla^2\Psi = 4\pi G (1+\mathcal{F})\bar\rho_{\rm m} \delta \, ,
\ee
where $\mathcal{F}$ can depend on time, mass and environment. Eq.~\eqref{eq:param_poisson} reduces to the standard Poisson equation for $\mathcal{F}=0$, and expressions for $\mathcal{F}$ in $f(R)$ and nDGP cosmologies are given in App.~\ref{sec:SC}. 

The mass fluctuation $\delta_{\rm i}$ at the initial time $a_{\rm i}$ within $R_{\smallsub TH}$ evolves as
\be\label{eq:SC_overdensity}
\delta = \left( \frac{R_{\rm i}}{R_{\smallsub TH}} \right)^{3}(1+\delta_{\rm i}) -1 \, ,
\ee
where $R_{\rm i}$ is the initial top-hat radius. Using Eqs.~\eqref{eq:SC1}-\eqref{eq:SC_overdensity} we then find $\delta_{\rm i}$ such that collapse (i.e. $R_{\smallsub TH}=0$) occurs at a chosen time $a=a_{\rm coll}$.
The Press-Schechter approach assumes that all regions in the initial density field with overdensities larger than $\delta_i$ have collapsed into halos by $a_{\rm coll}$. Equivalently, one can compare the linearly evolved initial fluctuations to the linearly extrapolated collapse overdensity $\delta_{\rm c}(a_{\rm coll}) \equiv D_\Lambda(a_{\rm coll})\delta_i/a_i$, with~\citep[see, e.g.,][]{Dodelson:2003}
\be
D_\Lambda(a) = \frac{5\Omega_{\rm m}}{2} \frac{H}{H_0} \int_0^a \frac{\diff a^\prime}{(a^\prime H/H_0)^3}
\ee
being the linear growth factor in $\Lambda$CDM\footnote{For cosmologies with a scale-independent linear growth, such as nDGP and $w$CDM, using the $\Lambda$CDM growth is simply a matter of convenience. In $f(R)$ gravity this approach has the advantage of preserving the statistics of the initial mass fluctuations. For more details see~\cite{Cataneo:2016}.}, and $\Omega_{\rm m} \equiv 8\pi G\bar\rho_{\rm m,0}/3H_0^2$.

In the idealised top-hat scenario, the spherical mass collapses to a point of infinite density. However, processes in the real Universe act so that, after turnaround, the overdensity eventually reaches virial equilibrium~\citep[see, e.g.,][]{Mo:2010}. Following~\cite{Schmidt:2010}~\citep[for an earlier work see also][]{Maor:2005}, we do not assume energy conservation during collapse, and compute the time of virialization, $a_{\rm vir}$, from the virial theorem alone (see App.~\ref{sec:SC} for details). This approach differs from previous works where changes induced by dark energy~\citep{Mead:2017} or modified gravity~\citep{Lombriser:2014} were neglected. The virial comoving radius $R_{\rm vir}$ of the formed halo can be derived from its virial mass
\be\label{eq:Mvir}
M_{\rm vir} = \frac{4\pi}{3}R_{\rm vir}^3 \, \bar\rho_{\rm m,0} \, \Delta_{\rm vir} \, ,
\ee
knowing that the virial overdensity is given by
\be\label{eq:vir_overdensity}
\Delta_{\rm vir} = [1+\delta(a_{\rm vir})] \left( \frac{a_{\rm coll}}{a_{\rm vir}} \right)^3 \, ,
\ee
with the mass fluctuation $\delta$ obtained from Eq.~\eqref{eq:SC_overdensity}.

\subsection{Halo model}\label{sec:hm}

The simplest statistics describing the clustering properties of the matter density field $\rho_{\rm m}({\bf x})$ is the 2-point correlation function or, its Fourier transform, the power spectrum $P(k)$ defined as
\be\label{eq:Pk_def}
\langle \tilde\delta({\bf k})\tilde\delta({\bf k^\prime}) \rangle \equiv (2\pi)^3 \delta_{\rm D}({\bf k} + {\bf k^\prime})P(k) \, ,
\ee
where $\delta_{\rm D}$ denotes the Dirac delta function, and $\tilde\delta({\bf k})$ represents the Fourier transform of the matter density fluctuations relative to the background mean density, $\delta({\bf x}) = \rho_{\rm m}({\bf x})/\bar\rho_{\rm m}-1$. Note that Eq.~\eqref{eq:Pk_def} assumes statistical homogeneity and isotropy.

In the halo model the matter power spectrum results from the contribution of correlations between halos ($P_{2\rm h}$) and those within halos ($P_{1\rm h}$), and can be written as\footnote{In this instance, and whenever the context is clear, we omit the time-dependence from our notation. However, we reintroduce it any time this can become a source of ambiguity.}
\be\label{eq:halo_model}
P(k) = P_{2\rm h}(k) + P_{1\rm h}(k) \, .
\ee
To properly account for these correlations we need to know the abundance of such halos. For any redshift $z$, the halo mass function provides the comoving number density of halos of mass $M_{\rm vir}$, and it is defined as
\be\label{eq:HMF_def}
n_{\rm vir} \equiv \frac{\diff n}{\diff \ln M_{\rm vir}} = \frac{\bar\rho_{\rm m,0}}{M_{\rm vir}}\nu f(\nu) \frac{\diff \ln \nu}{\diff \ln M_{\rm vir}} \, ,
\ee
where the peak height $\nu \equiv \delta_{\rm c}/\sigma$, and we adopt the Sheth-Tormen (ST) multiplicity function~\citep{Sheth:1999,Sheth:2002}
\be\label{eq:SThmf}
\nu f(\nu) = A\sqrt{\frac{2}{\pi}q\nu^2} \left[1 + \left( q\nu^2 \right)^{-p} \right]\exp\left[ -q\nu^2/2 \right] \, .
\ee
Here, the normalization constant $A$ is found imposing that all mass in the Universe is confined into halos, i.e. $\int \diff\nu f(\nu) = 1$, and the remaining parameters take the $\Lambda$CDM standard values $q=0.75$ and $p=0.3$, unless stated otherwise. The variance of the linear density field smoothed with a top-hat filter of comoving radius $R$ enclosing a mass $M=4\pi R^3\bar\rho_{\rm m,0}/3$ is given by
\be\label{eq:variance}
\sigma^2(R,z) = \int \frac{\diff^3 k}{(2\pi)^3} |\tilde W(kR)|^2 P_{\rm\scriptscriptstyle L}(k,z) \, ,
\ee
where $\tilde W$ is the Fourier transform of the top-hat filter, and $P_{\rm\scriptscriptstyle L}(k,z)$ is the $\Lambda$CDM linear power spectrum. At this point it is worth emphasising that in some GR extensions, besides its usual dependence on background cosmology and redshift, the spherical collapse threshold $\delta_{\rm c}$ can also vary with halo mass and environment~\citep{Li:2012,Li:2012xxx,Lam:2012xxx,Lombriser:2013,Lombriser:2014}. When appropriate we include both these dependencies in our modelling by following the approach of~\cite{Cataneo:2016}, where the initial value of the environmental overdensity is derived from the peak of the environment probability distribution.

Halos are biased tracers of the underlying dark matter density field, and at the linear level the halo and matter density fields are connected by the relation $\delta_{\rm h} = b_{\rm\scriptscriptstyle L} \delta$. Adopting the ST mass function, the peak-background split formalism predicts the linear halo bias\footnote{\cite{Valogiannis:2019} recently found that in $f(R)$ gravity the linear halo bias contains an additional term accounting for the environmental dependence, which we omit in Eq.~\eqref{eq:bias}. Given the relative unimportance of the bias for our halo model reactions (see Secs.~\ref{sec:reactions} and~\ref{sec:results}), this choice is, in effect, inconsequential for the accuracy of our predictions.}~\citep{Sheth:1999}
\be\label{eq:bias}
b_{\rm\scriptscriptstyle L}(M_{\rm vir}) = 1 + \frac{q\nu^2 - 1}{\delta_{\rm c}} + \frac{2p}{\delta_{\rm c}[1+(q\nu^2)^p]} \, .
\ee

The last piece of information required by the halo model is a description of the matter distribution within halos. We adopt Navarro-Frenk-White (NFW) halo profiles~\citep{Navarro:1996} \\
\be\label{eq:NFW}
\rho_{\rm h}(r) = \frac{\rho_{\rm s}}{r/r_{\rm s}(1+r/r_{\rm s})^2} \, ,
\ee
where the scale radius $r_{\rm s}$ is parametrized through the virial concentration $c_{\rm vir} \equiv R_{\rm vir}/r_{\rm s}$, and the normalization $\rho_{\rm s}$ follows from the virial mass as
\be\label{eq:rho_s}
\rho_{\rm s} = \frac{M_{\rm vir}}{4\pi r_{\rm s}^3} \left[ \ln(1+c_{\rm vir}) - \frac{c_{\rm vir}}{1+c_{\rm vir}} \right]^{-1} \, .
\ee
Inside the virial radius, and for all cosmological models studied here, the NFW profiles are a good representation of the averaged halo profiles measured in simulations~\citep{Schmidt:2009c,Schmidt:2009a,Zhao:2011,Lombriser:2012,Kwan:2013,Shi:2015aya,Achitouv:2016}.

In $\Lambda$CDM, $f(R)$ gravity and nDGP we model the $c$-$M$ relation as the power law
\be\label{eq:concentration}
c_{\rm vir}(M_{\rm vir},z) = \frac{c_0}{1+z} \left( \frac{M_{\rm vir}}{M_\ast} \right)^{-\alpha} \, ,
\ee
fixing $c_0=9$ and $\alpha=0.13$~\citep{Bullock:2001}, and $M_\ast$ is defined by $\nu(M_\ast)=1$. In particular, for $f(R)$ gravity $M_\ast$ depends itself on the halo mass~\citep{Lombriser:2014}, which means the $c$-$M$ relation for these models is no longer described by a simple power law~\citep{Shi:2015aya}. For the smooth dark energy models in Sec.~\ref{sec:DE} we correct for the different expansion histories following~\cite{Dolag:2004}, that is
\be\label{eq:concentration_de}
c_{\rm vir} \rightarrow \frac{c_0}{1+z} \left( \frac{M_{\rm vir}}{M_\ast} \right)^{-\alpha} \frac{g_{\smallsub DE}(z \rightarrow \infty)}{g_\Lambda(z \rightarrow \infty)} \, ,
\ee
where $g_{\smallsub X}$ is the linear growth factor normalized to $z=0$ (see App.~\ref{sec:PT}). This correction reflects that halos collapse at different times in cosmological models with different growth histories. In cosmological models where halos collapse earlier these halos will be more concentrated compared to the same mass halos if they form later. In App.~\ref{sec:tests} we demonstrate that our results are insensitive to the correct shape of the $c$-$M$ relation on scales $k \lesssim 0.5$ \hMpc.

We can now predict the nonlinear matter power spectrum, and rewrite Eq.~\eqref{eq:halo_model} as
\be
P(k) = I^2(k)P_{\scriptscriptstyle \rm L}(k) + P_{1\rm h}(k) \, ,
\ee
where, more explicitly,
\be
P_{1\rm h}(k) = \int \diff\ln M_{\rm vir} \, n_{\rm vir} \left( \frac{M_{\rm vir}}{\bar\rho_{\rm m,0}} \right)^2 |u(k,M_{\rm vir})|^2,\label{eq:1h_integral} \label{eq:P1h} \\
I(k) = \int \diff\ln M_{\rm vir} \, n_{\rm vir} \frac{M_{\rm vir}}{\bar\rho_{\rm m,0}}u(k,M_{\rm vir}) b_{\rm\scriptscriptstyle L}(M_{\rm vir}) \, . \label{eq:2h_integral}
\ee
In the equations above, $u(k,M)$ corresponds to the Fourier transform of an NFW profile truncated at $R_{\rm vir}$, normalized such that $u(k \rightarrow 0,M) \rightarrow 1$. Note that from Eqs.~\eqref{eq:SThmf} and~\eqref{eq:bias} it follows that $\lim_{k \rightarrow 0} I(k) = 1$.

\subsection{Halo model reactions}\label{sec:reactions}

The apparent simplicity and versatility of the halo model has contributed to its widespread use as a method to predict the nonlinear matter power spectrum in diverse scenarios. Examples include the $\Lambda$CDM cosmology~\citep{Seljak:2000,Peacock:2000,Giocoli:2010,Valageas:2011,Valageas:2013,Mohammed:2014,Seljak:2015,Daalen:2015,Mead:2015,Schmidt:2016}, dark energy and modified gravity models~\citep{Schmidt:2009c,Schmidt:2010,Li:2011,Fedeli:2012,Brax:2013,Lombriser:2014,Barreira:2014a,Barreira:2014b,Achitouv:2016,Mead:2016,Hu:2018}, massive neutrinos~\citep{Abazajian:2005,Massara:2014,Mead:2016}, baryonic physics~\citep{Mohammed:2014,Fedeli:2014a,Fedeli:2014b,Mead:2015}, alternatives to cold dark matter~\citep{Dunstan:2011,Schneider:2012,Marsh:2016}, and primordial non-Gaussianity~\citep{Smith:2011}. Its imperfect underlying assumptions are however responsible for inaccuracies that limit its applicability to future high-quality data~\citep[see, e.g., Figure 1 in][]{Massara:2014}, where percent level accuracy is required in order to obtain unbiased cosmological constraints~\citep{Huterer:2005,Eifler:2011,Hearin:2012,Taylor:2018}.

To mitigate these downsides one can add complexity to the model at the expense of introducing new free parameters~\citep[see, e.g.,][]{Seljak:2015}, fitting the existing ones to the matter power spectrum measured in simulations~\citep[see, e.g.,][]{Mead:2015}, or sensibly increasing the computational costs by going beyond linear order in perturbation theory~\citep[see, e.g.,][]{Valageas:2011}. Here, instead, we follow and extend the approach presented in~\cite{Mead:2017}, which we shall refer to as \emph{halo model reactions}\footnote{Note that in~\cite{Mead:2017} this is referred to as \emph{response}. We use the term \emph{reaction} to distinguish it from the quantities studied in~\cite{Neyrinck:2013},~\cite{Nishimichi:2016} or~\cite{Barreira:2017}. Our and these other definitions are all conceptually analogous, in the sense that they describe how the nonlinear power spectrum responds to changes in some feature, which in our case is physics beyond the vanilla $\Lambda$CDM cosmology, e.g. fifth forces, evolving dark energy, massive neutrinos, baryons etc.}. 

Our goal is to model the nonlinear power spectrum of fairly general extensions to the standard cosmology, a flat $\Lambda$CDM Universe with massless neutrinos. These cosmologies equipped with beyond-$\Lambda$CDM physics are what we will call \emph{real} cosmologies. We use the halo model to determine the change (i.e. the reaction) that this new physics induces in a reference $\Lambda$CDM cosmology, for which simulations are considerably cheaper. Key to the success of our method is how this reference cosmology is defined, which is what we will call the \emph{pseudo} cosmology. Essentially, this is a $\Lambda$CDM cosmology evolved with standard gravity up to a final redshift $z_{\rm f}$, with the additional property that its linear clustering of matter exactly matches that of the target \emph{real} cosmology of interest at $z_{\rm f}$.
In other words, the cold dark matter and the cosmological constant determine the expansion history and growth of structure of the \emph{pseudo} cosmology, but the initial conditions (see Sec.~\ref{sec:sims}) are adjusted so that
\be\label{eq:Plin_equality}
P_{\rm\scriptscriptstyle L}^{\rm pseudo}(k,z_{\rm f}) = P_{\rm\scriptscriptstyle L}^{\rm real}(k,z_{\rm f}) \, .
\ee
The reaction function is then defined as the ratio of the nonlinear matter power spectrum in the \emph{real} cosmology to that in the \emph{pseudo} cosmology, 
\be\label{eq:resp_def}
\mathcal{R}(k,z) \equiv \frac{P^{\rm real}(k,z)}{P^{\rm pseudo}(k,z)} \, ,
\ee
and our corresponding halo model prediction takes the heuristic form\footnote{Note that we neglect the integral factor Eq.~\eqref{eq:2h_integral} in our two-halo terms. We checked that setting $I^2(k) = 1$ for all scales has no measurable impact on our halo model reactions.}
\be\label{eq:resp_hm}
\mathcal{R}(k,z) = \frac{[(1-\mathcal{E})e^{-k/k_\star}+\mathcal{E}]P_{\rm\scriptscriptstyle L}^{\rm real}(k,z) + P_{1\rm h}^{\rm real}(k,z)}{P_{\rm\scriptscriptstyle L}^{\rm real}(k,z) + P_{1\rm h}^{\rm pseudo}(k,z)} \, .
\ee
Here, $\mathcal{E} \sim 1$ and $k_\star > 0$ are parameters introduced to improve the accuracy of the halo model reactions in modified gravity theories. These are not free parameters, and we shall derive them using the halo model and standard perturbation theory below. However, let us first examine the general behaviour of Eq.~\eqref{eq:resp_hm}:
\begin{enumerate}
\item On large linear scales $\mathcal{R} \rightarrow 1$ by definition\footnote{This is not strictly true in the traditional halo model implementation we adopt in this work. In fact, the one-halo terms have a constant tail in the low-$k$ limit (see Eq.~\ref{eq:1h_constant}) that dominates the two-halo contributions on very large scales. In a consistent formulation of the halo model, however, where mass and momentum conservation are enforced, this tail disappears leaving only the theoretically motivated two-halo term~\citep{Schmidt:2016}. For our purposes we can simply ignore this inconsistency, and restrict the use of the reaction function to scales $k > 0.01$ \hMpc.};
\item On small nonlinear scales $\mathcal{R} \approx P_{1\rm h}^{\rm real}/P_{1\rm h}^{\rm pseudo}$;
\item Quasi-linear scales $0.01 \lesssim k \, \Mpch \lesssim 0.1$ are well described by perturbation theory, while intermediate scales $0.1 \lesssim k \, \Mpch \lesssim 1$ are primarily controlled by the halo mass function ratio $n_{\rm vir}^{\rm real}/n_{\rm vir}^{\rm pseudo}$.
\end{enumerate}
Fixing the \emph{real} and \emph{pseudo} linear power spectra to be identical (as in Eq.~\ref{eq:Plin_equality}) forces the corresponding mass functions to be somewhat similar. Therefore, owing to (iii), reaction functions can overcome the typical inaccuracies that plague the halo model in the transition region between large and small scales.

To assign a value to the boost/suppression term $\mathcal{E}$, it is important to realise that we would like to preserve the smoothness in the transition from the linear to the nonlinear regime. In turn, this is tied to the shape of the one-halo terms on scales $0.5 \lesssim k \, \Mpch \lesssim 1$, where the two-halo contribution becomes subdominant. In this regime the one-halo terms are well approximated by their large-scale limit $P_{1\rm h}(k \rightarrow 0)$, thus suggesting that
\be\label{eq:1h_constant}
\mathcal{E}(z) = \frac{P_{1\rm h}^{\rm real}(k \rightarrow 0,z)}{P_{1\rm h}^{\rm pseudo}(k \rightarrow 0,z)}
\ee 
is a good choice, one that only depends on the ratio $n_{\rm vir}^{\rm real}/n_{\rm vir}^{\rm pseudo}$.

In Eq.~\eqref{eq:resp_hm}, the transition rate from linear to nonlinear scales is governed by the parameter $k_\star~$\footnote{As we shall see below, $k_\star$ is derived from perturbation theory and is largely independent of the one-halo contribution. However, its specific value should be interpreted with caution, in that equally good alternatives to the exponential function in Eq.~\eqref{eq:resp_hm} can provide different solutions to Eq.~\eqref{eq:kstar}.}: for $k_\star \rightarrow 0$ the halo model reaction collapses to the ratio of one-halo terms; in the opposite case, $k_\star \rightarrow \infty$, the parameter $\mathcal{E}$ loses any role, and the reaction reduces to its definition in~\cite{Mead:2017}. We determine this scale using standard perturbation theory (SPT)~\citep{Koyama:2009,Brax:2013,Bose:2016} by solving the equation
\be\label{eq:kstar}
\mathcal{R}(k_0,z | k_\star) = \frac{P_{\text{SPT}}^{\rm real}(k_0,z) + P_{1\rm h}^{\rm real}(k_0,z)}{P_{\text{SPT}}^{\rm pseudo}(k_0,z) + P_{1\rm h}^{\rm pseudo}(k_0,z)} \, ,
\ee
where
\be\label{eq:P1loop}
P_{\text{SPT}}(k,z) = P_{\rm \scriptscriptstyle L}(k,z) + P_{22}(k,z) + P_{13}(k,z) + P_{13}^\Psi(k,z) \, ,
\ee
and we set $k_0 = 0.06$ \hMpc, which we found to be the largest wavenumber that can both ensure reliable perturbative predictions and keep the inaccuracies induced by the exponential sensitivity to $k_\star$ under control~\citep[see App.~\ref{sec:PT} and][]{Carlson:2009}. Expressions for the second order corrections $P_{22}$, $P_{13}$ and $P_{13}^\Psi$ to the linear power spectrum are given in App.~\ref{sec:PT}. Note that alternative perturbation schemes can also be used in Eq.~\eqref{eq:kstar}, such as the Lagrangian Perturbation Theory for modified gravity recently developed in~\cite{Aviles:2017}.

The role of the two-halo correction factor in Eq.~\eqref{eq:resp_hm} becomes clear in the limiting case $k_\star \rightarrow \infty$, where it goes to unity. \cite{Mead:2017} showed that this form of the reactions matches smooth dark energy simulations at percent level or better on scales $k \lesssim 1$ \hMpc for $z=0$ (see also Sec.~\ref{sec:results} below). On quasi-linear scales this remarkable agreement can be understood in terms of standard perturbation theory. Schematically, the nonlinear matter power spectrum is a non-trivial function of the linear power spectrum obtained through some operator $\mathcal{K}$, i.e. $P(k) = \mathcal{K}[P_{\rm \scriptscriptstyle L}(k)]$~\citep{Bernardeau:2002}. Provided that gravitational forces remain unchanged, then Eq.~\eqref{eq:Plin_equality} enforces $P^{\rm real} \approx P^{\rm pseudo}$ on linear and quasi-linear scales. This is no longer true for modifications of gravity, since the structure of the $\mathcal{K}_{\rm MG}$ operator is altered by different mode-couplings and screening mechanisms. We correct for this fact by including the two-halo pre-factor in Eq.~\eqref{eq:resp_hm}, so that finite $k_\star$ roughly encapsulates the extent of the mismatch between $\mathcal{K}_{\rm MG}$ and $\mathcal{K}_{\rm GR}$. For comparison, in F5 at $z=0$ we have $k_\star = 0.3$ \hMpc, whereas in nDGPm it becomes $k_\star = 0.95$ \hMpc, which reflects the different screening efficiency on large scales between the chameleon and Vainshtein mechanisms.

Although our choice of $k_0$ is commonly regarded as well within the quasi-linear regime, screening mechanisms in modified gravity induce nonlinearities on large scales that can be more important than in GR. For this, the determination of $k_\star$ can be complicated by inaccuracies specific to the perturbation theory employed, and to reduce their impact on the halo model reactions we take advantage of the following two facts: (i) on large scales we expect $P_{\rm NoScr}^{\rm real} \approx P^{\rm pseudo}$, where $P_{\rm NoScr}^{\rm real}$ denotes the nonlinear matter power spectrum of the \emph{real} cosmology assuming there is no screening mechanism; (ii) the kernel operators $\mathcal{K}_{\rm MG}^{\rm Scr}$ and $\mathcal{K}_{\rm MG}^{\rm NoScr}$ for the screened and unscreened \emph{real} cosmology, respectively, have a similar structure (see App.~\ref{sec:PT}). Therefore, at least in principle, the ratio $P_{\rm Scr,SPT}^{\rm real}/P_{\rm NoScr,SPT}^{\rm real}$ could give a better description of the reaction on large scales than the obvious candidate $P_{\rm Scr,SPT}^{\rm real}/P_{\rm SPT}^{\rm pseudo}$. Hereafter we will use $P_{\rm NoScr,SPT}^{\rm real}$ instead of $P_{\rm SPT}^{\rm pseudo}$ in Eq.~\eqref{eq:kstar}, which in spite of being a sub-optimal strategy in some cases (see right panel of Figure~\ref{fig:SPTreacFoR}) produces the most consistent behaviour across the cosmological models we have tested, as shown in Sec.~\ref{sec:results}.

In summary, halo model reactions provide a fast (we only need one-loop SPT for a single wavenumber) and general framework to map accurate nonlinear matter power spectra in $\Lambda$CDM to other non-standard cosmologies. We apply this method to $f(R)$ gravity, nDGP and evolving dark energy, and test its performance in Sec.~\ref{sec:results} against the cosmological simulations described in the next section.

\section{Simulations}
\label{sec:sims} 
\begin{table*}
\centering
\caption{Main technical details of the simulations employed in this work. The Nyquist frequency $k_{\rm Ny}=\pi N_p/L_{\rm box}$, $\epsilon$ is the force resolution, and $m_p$ the particle mass. The standard cosmological parameters for the \emph{real} $f(R)$ and nDGP simulations, as well as for their standard $\Lambda$CDM counterparts, are $\omega_{\rm b} \equiv \Omega_{\rm b}h^2 = 0.02225$, $\omega_c \equiv \Omega_{\rm c}h^2  = 0.1198$, $H_0 = 100h = 68$ km s$^{-1}$ Mpc$^{-1}$, $A_{\rm s} = 2.085 \times 10^{-9}$, $n_{\rm s} = 0.9645$. For the evolving dark energy models we have instead $\omega_{\rm b} = 0.0245$, $\omega_{\rm c} = 0.1225$, $H_0 =70$ km s$^{-1}$ Mpc$^{-1}$, $\sigma_8 = 0.8$, $n_{\rm s} = 0.96$.}
\label{tab:sims}
\begin{tabular}{|c|c|c|c|c|c|c|c|}
\hline\hline
model 	& $L_{\rm box}$ & $N_p^3$ & $k_{\rm Ny}$ & $\epsilon$ & $m_p$  & realisations & code \\ \hline
 $f(R)$     	& 512 $\Mpch$ & $1024^3$ & 6.3 \hMpc & 15.6 kpc$\, h^{-1}$ & $1.1 \times 10^{10} \, M_\odot h^{-1}$ & 1 & {\sc ecosmog} \\ \hline
  nDGP    	& 512 $\Mpch$ & $1024^3$ & 6.3 \hMpc & 15.6 kpc$\, h^{-1}$ & $1.1 \times 10^{10} \, M_\odot h^{-1}$ & 1 & {\sc ecosmog} \\ \hline
  DE  	& 200 $\Mpch$ & $512^3$ & 8 \hMpc & 7.8 kpc$\, h^{-1}$ & $5 \times 10^{9} \, M_\odot h^{-1}$ & 3 & {\sc gadget}-2 \\ \hline\hline
\end{tabular}
\end{table*}

The simulations of $f(R)$ gravity and DGP models used in this work were run using {\sc ecosmog} \citep{Li:2011vk,Li:2013nua,Li:2013tda}, which has been developed to simulate the structure formation in various subclasses of models within the Horndeski family of theories. {\sc ecosmog} is an extension of the simulation code {\sc ramses} \citep{Teyssier:2001cp}, which is a particle-mesh code employing adaptive mesh refinement to achieve high force resolution. The simulations are dark matter only and run in boxes with comoving size $512 \, h^{-1}$Mpc using $1024^3$ simulation particles. Other basic information can be found in Table~\ref{tab:sims}. The initial conditions of the simulations are generated using 2{\sc lpt}ic \citep{Crocce:2006ve}, which calculates the particle initial displacements and peculiar velocities up to second order in Lagrangian perturbations, allowing us to start from a relative low initial redshift $z_{\rm ini}=49$. To isolate the effect of nonlinearities we use identical phases for the initial density field in all cases. The linear power spectra used to generate the initial conditions are computed using {\sc camb}~\citep{Lewis:2000}, with $\Omega_{\rm m}=0.3072$, $\Omega_\Lambda=0.6928$, $h=0.68$, $\Omega_{\rm b}=0.0481$ for all simulations. Importantly, following Eq.~(\ref{eq:Plin_equality}) the normalisation -- and shape in $f(R)$ gravity -- of the initial linear power spectra are different in the \emph{real} and \emph{pseudo} simulations. Since at early times deviations from GR are negligible, simulations in $\Lambda$CDM and modified gravity share the same initial conditions set by the $\Lambda$CDM power spectrum,
	\be\label{eq:ini_real_mg}
	P_{\rm\smallsub L}^{\rm real}(k,z_{\rm ini}) = \left[ \frac{D_\Lambda(z_{\rm ini})}{D_\Lambda(z=0)} \right]^2 P_{\rm\smallsub L}^\Lambda(k,z=0) \, ,
	\ee
	with $\sigma_8(z=0)=0.8205$. 
The \emph{pseudo} runs (to which we will apply the halo model reactions) have different initial conditions, generated using modified gravity linear power spectra at the final redshift, $z_{\rm f}$, and then rescaled with the $\Lambda$CDM linear growth to the starting redshift as
	\be\label{eq:ini_pseudo_mg}
	P_{\rm\smallsub L}^{\rm pseudo}(k,z_{\rm ini}) = \left[ \frac{D_\Lambda(z_{\rm ini})}{D_\Lambda(z_{\rm f})} \right]^2 P_{\rm\smallsub L}^{\rm MG}(k,z_{\rm f}) \, ,
	\ee
where in this work $z_{\rm f}=0$ or 1. By evolving the initial \emph{real} and \emph{pseudo} power spectra, Eqs.~\eqref{eq:ini_real_mg} and~\eqref{eq:ini_pseudo_mg}, with the modified and standard laws of gravity, respectively, Eq.~\eqref{eq:Plin_equality} will be automatically satisfied at $z_{\rm f}$. We extract the nonlinear matter power spectrum from our particle snapshots using the public code {\sc powmes}~\citep{Colombi:2009}.

Simulations of dark energy models were run using a modified version of {G{\sc adget}}-2 that allows for the $\{ w_0,w_a \}$ parametrisation under the assumption that the dark energy is homogeneous. Initial conditions were generated at $z_{\rm ini} = 199$ using N-G\textsc{en}IC~\citep{Springel:2015}, a code that calculates initial particle displacements and peculiar velocities based on the Zeldovich approximation. Our simulations take place in 200 $\Mpch$ boxes and use $512^3$ particles. Note that since we are concerned only with \emph{ratios} of power spectra the overall resolution requirements on the simulations are less stringent than if we were interested in the \emph{absolute} power spectra. We checked that our simulated reactions were insensitive to the realisation, box size, particle number and softening up to the wave numbers we show. 

Differently from the modified gravity runs, we fix $\sigma_8 = 0.8$ for all the evolving dark energy models. Then, for the \emph{real} cosmologies the amplitudes of the initial density field are determined by
	\be\label{eq:ini_real_de}
	P_{\rm\smallsub L}^{\rm real}(k,z_{\rm ini}) = \left[ \frac{D_{\rm DE}(z_{\rm ini})}{D_{\rm DE}(z=0)} \right]^2 P_{\rm\smallsub L}^{\rm DE}(k,z=0) \, ,
	\ee
	where $D_{\rm DE}(z)$ is the linear growth of a specific dark energy model, while for the \emph{pseudo} counterparts one simply replaces $P_{\rm\smallsub L}^{\rm MG}$ with $P_{\rm\smallsub L}^{\rm DE}$ in Eq.~\eqref{eq:ini_pseudo_mg}.

\section{Results}\label{sec:results}

Here we test our halo model reactions (see Sec.~\ref{sec:reactions}) against the same quantities constructed from the \emph{real} and \emph{pseudo} cosmological simulations described in Sec.~\ref{sec:sims}. To help get a better sense of the performance of our method, for each \emph{real} cosmology $\mathcal{C}$ we also compute the standard ratios $P_{\mathcal{C}}^{\rm real}(k)/P_{\Lambda{\rm CDM}}(k)$, where our theoretical prediction for the \emph{real} nonlinear power spectrum is obtained as
\be\label{eq:pseudo_to_real}
P_{\mathcal{C}}^{\rm real}(k) = \mathcal{R}(k) \times P_{\mathcal{C}}^{\rm pseudo}(k) \, ,
\ee
where $\mathcal{R}(k)$ is given by Eq.~\eqref{eq:resp_hm}. To test our modified gravity predictions we calculate $P_{\mathcal{C}}^{\rm pseudo}(k)$ from HM{\sc code}~\citep{Mead:2015,Mead:2016} or using the measurement from the simulations directly. For evolving dark energy, instead, we use the \emph{pseudo} nonlinear power spectrum given by the Coyote Universe emulator~\citep{Heitmann:2014}, and in so doing we illustrate how one could predict the \emph{real} power spectrum for cosmologies beyond the concordance model with the aid of a carefully designed $\Lambda$CDM-\emph{like} emulator (Giblin et al. in prep.).

\subsection{$f(R)$ gravity}

\begin{figure*}
\begin{center}
\includegraphics[width=\columnwidth]{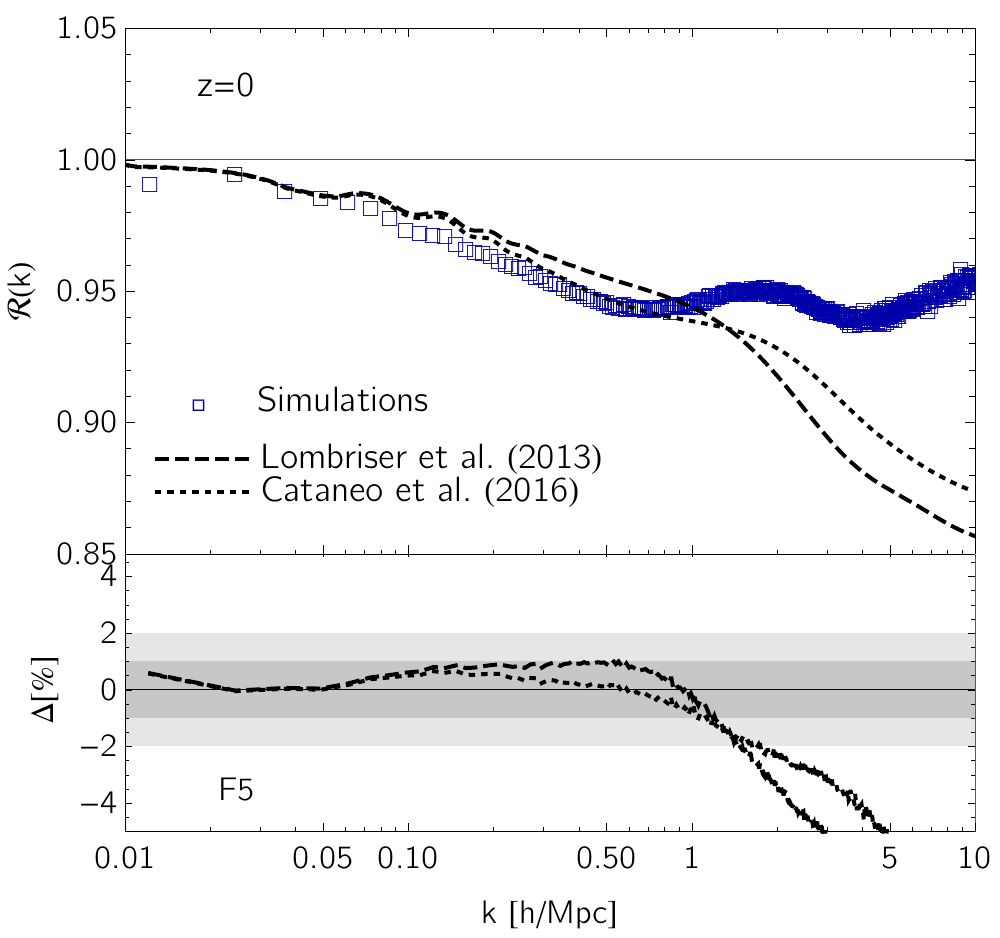}
\quad
\includegraphics[width=\columnwidth]{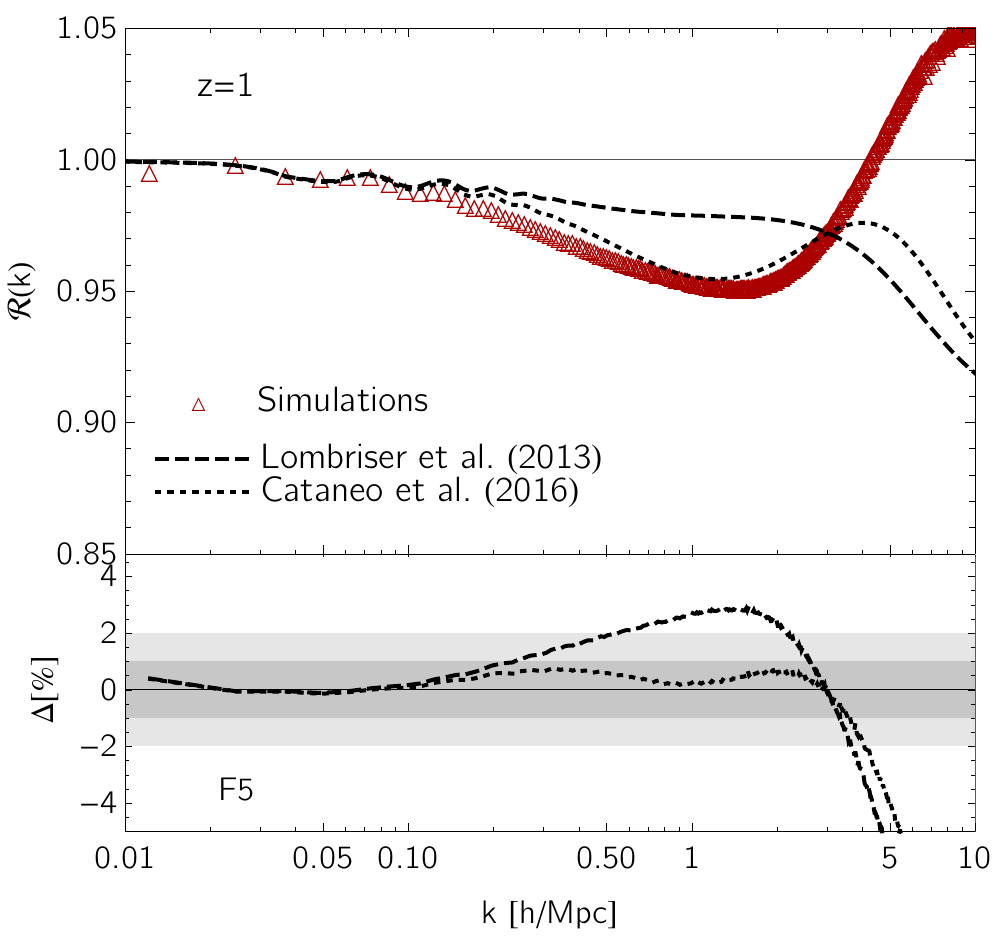}
\end{center}
\caption{Matter power spectrum reactions in $f(R)$ gravity for $|\fRbg|=10^{-5}$. In each panel the data points represent the reactions measured from simulations; lines denote the corresponding halo model predictions identified by the halo mass function used for the one-halo contributions in $f(R)$ gravity,  that is either \citet{Lombriser:2013} (dashed) or \citet{Cataneo:2016} fits (dotted). Lower panels present the fractional deviation of the halo model relative to the simulations, with grey bands marking the $1\%$ and $2\%$ uncertainty regions. {\it Left:} reaction at $z=0$ with both halo mass functions providing predictions within $1\%$ from the simulations for $k \lesssim 1$ \hMpc, as shown in the lower panel.  {\it Right:} $z=1$ reaction. The lower panel shows that thanks to the improved semi-analytical prescription for the halo abundances in~\citet{Cataneo:2016} the agreement between halo model and simulations reaches percent-level on scales $k \lesssim 1$ \hMpc.}
\label{fig:respF5}
\end{figure*}

\begin{figure*}
\begin{center}
\includegraphics[width=\columnwidth]{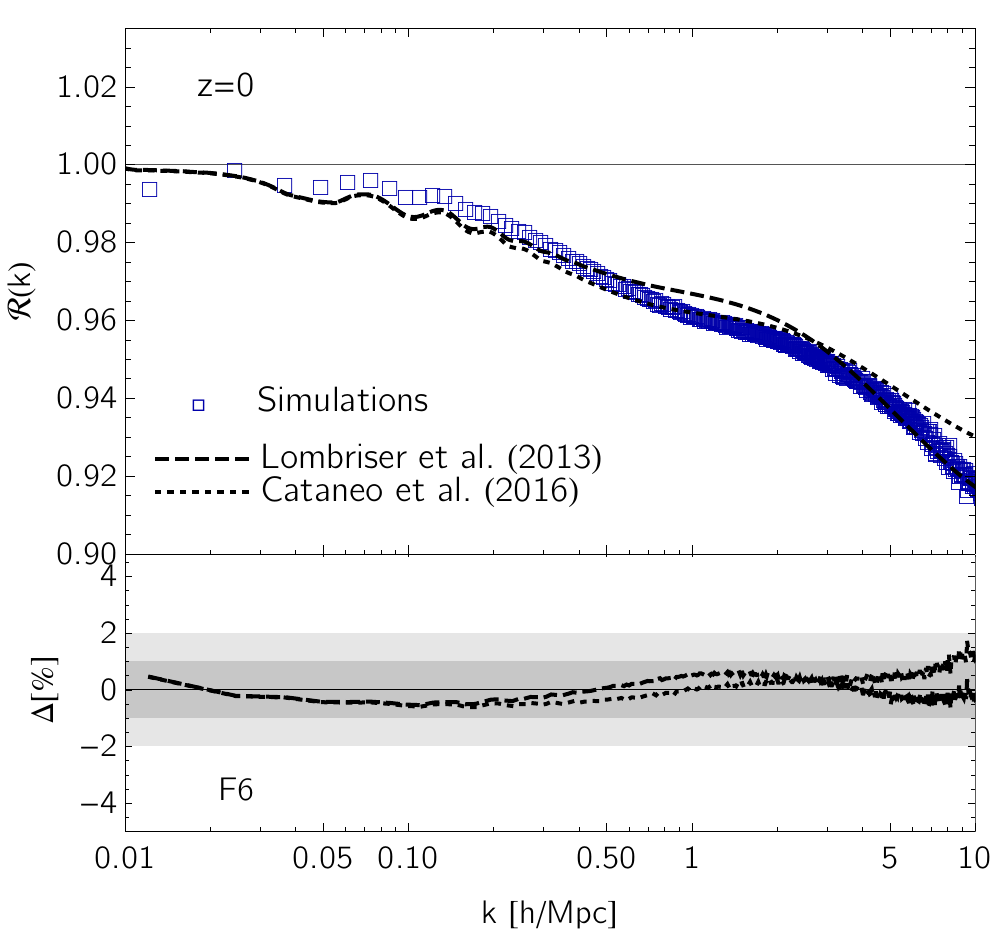}
\quad
\includegraphics[width=\columnwidth]{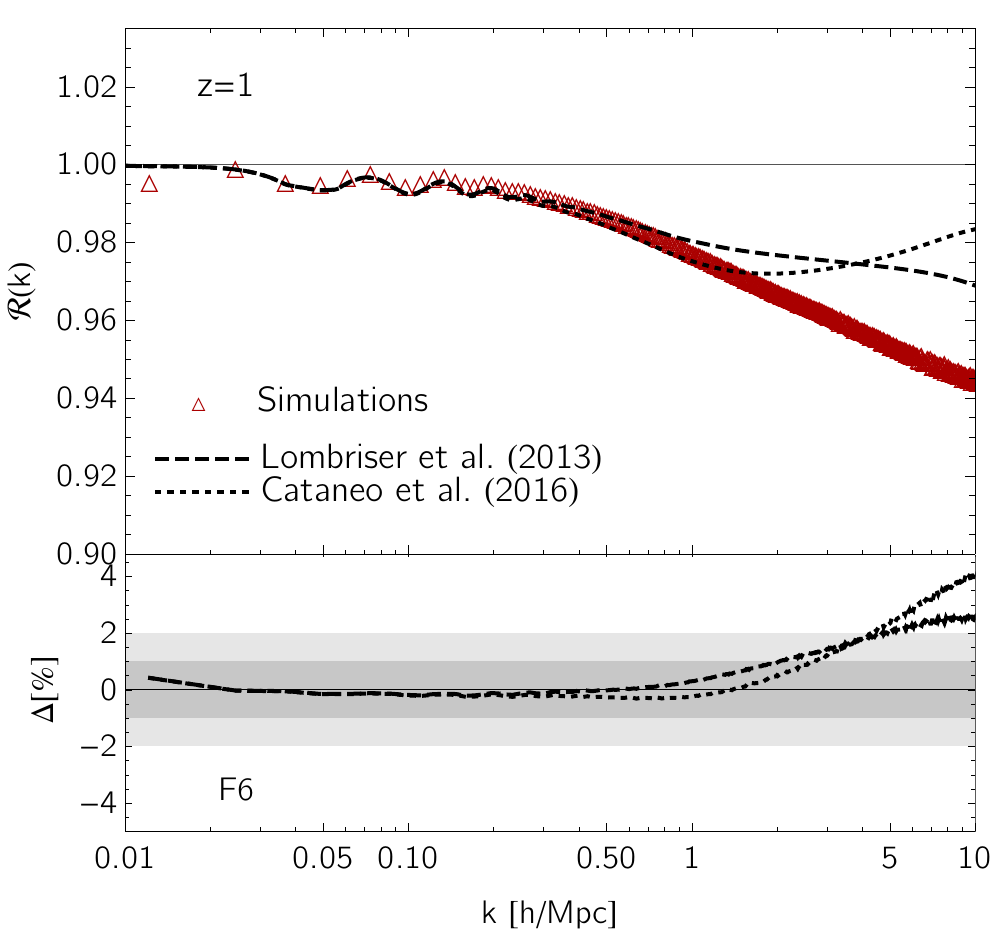} 
\end{center}
\caption{Same as Figure \ref{fig:respF5} with the amplitude of the background scalaron field now fixed to $|\fRbg|=10^{-6}$. For this cosmology both halo mass functions perform very well regardless of redshift, which can be explained by their similarity as shown in \citet{Cataneo:2016}.}
\label{fig:respF6}
\end{figure*}

Figure~\ref{fig:respF5} shows the matter power spectrum reactions calculated using Eq.~\eqref{eq:resp_hm} for the F5 cosmology at $z=0$ (left panel) and $z=1$ (right panel) in comparison to the measured reactions in the $N$-body simulations. For the virial halo mass function entering $P_{1\rm h}$ (see Eq.~\ref{eq:1h_integral}) we first adopt the approach developed in~\cite{Lombriser:2013}, which incorporates both self- and environmental-screening in the spherical collapse (see App.~\ref{sec:SC}). At $z=0$ the halo concentrations, virial radii and mass functions are good enough to give percent level predictions on scales $k \lesssim 1$ \hMpc. A deviation of a few percent is however visible at $z=1$ starting on scales as large as $k \approx 0.2$ \hMpc. In App.~\ref{sec:tests} we show that changes in the halo profiles only affect scales $k \gtrsim 0.5$ \hMpc, suggesting that the observed inaccuracies could be caused by a  mismatch between the predicted virial mass function ratio $n_{\rm vir}^{\rm real}/n_{\rm vir}^{\rm pseudo}$ and the same quantity measured in simulations. Indeed, \cite{Cataneo:2016} found that, for halo masses defined by spherical regions with an average matter density 300 times the mean matter density of the Universe, the halo mass function of~\cite{Lombriser:2013} can deviate up to 10\% from the simulations. 

Given the complexity of measuring the virial halo mass function in $f(R)$ simulations\footnote{Due to the nature of the chameleon screening, in $f(R)$ gravity the virial overdensity depends on both the mass of the halo and the gravitational potential in its environment. Things are much simpler in DGP, where by virtue of the Vainshtein screening both dependencies disappear.}, we investigate changes in the reactions induced by a more accurate description of the halo abundances with the fits provided in~\cite{Cataneo:2016}. There, however, the calibration of the $n_\Delta^{f(R)}/n_\Delta^{\rm \Lambda CDM}$ ratios was performed for $\Delta = 300$, whereas for our purposes we need $\Delta = \Delta_{\rm vir}$. We go from one mass definition to the other with the scaling relations outlined in~\cite{Hu:2003}~\citep[for a first application to $f(R)$ gravity see][]{Schmidt:2009c}. Inevitably, this transformation suffers from inaccuracies in $c_{\rm vir}$ and $\Delta_{\rm vir}$, which we attempt to compensate for by adjusting the $M_{300}(M_{\rm vir})$ relation so that the new rescaled mass function provides a present-day halo model reaction that is at least as good as the reaction obtained when using the~\cite{Lombriser:2013} virial mass function (dotted line in the left panel of Figure~\ref{fig:respF5})\footnote{In practice, we start with the~\cite{Hu:2003} relation $M_{300} = \mathcal{Q}(M_{\rm vir})M_{\rm vir}$, and make the replacement $\mathcal{Q}(M_{\rm vir}) \rightarrow \mathcal{Q}^{\prime}(M_{\rm vir}) = \min[{\tilde a}\mathcal{Q}^{f(R)}({\tilde b}M_{\rm vir}),\mathcal{Q}^{\rm GR}(M_{\rm vir})]$, where ${\tilde a}$ and ${\tilde b}$ are $\mathcal{O}(1)$ free parameters fine-tuned to reach the required accuracy in the halo model reaction at $z=0$. The use of the minimum operator ensures that the $f(R)$ conversion factor matches the corresponding GR value for masses large enough to fully activate the chameleon screening.}. We then take the ratio of the~\cite{Cataneo:2016} rescaled mass function to that of~\cite{Lombriser:2013}, and treat this quantity as a correction factor for the latter. To find the required adjustment at high redshifts we shift the $z=0$ correction by an amount $\Delta\log_{10} M_{\rm vir}$ inferred from the redshift evolution of the ratio of the two halo mass functions over the range $z \in [0,0.5]$~\citep[see central panel of Figure 4 in][]{Cataneo:2016}. A simple extrapolation to $z=1$ gives $\Delta\log_{10} M_{\rm vir} = 1$. Although far from being a rigorous transformation, the resulting halo model reaction now agrees to better than 1\% down to $k \approx 3$ \hMpc, as shown in the right panel of Figure~\ref{fig:respF5}. Figure~\ref{fig:respF6} illustrates that similar considerations are also valid for the F6 cosmology, where we use the same mass shift $\Delta\log_{10} M_{\rm vir} = 1$ to go from the $z=0$ to the $z=1$ mass function correction. In all cases, deviations in the highly nonlinear regime are most likely caused by inaccurate $c$-$M$ relations. We leave the study of $f(R)$ gravity reactions on small scales derived from proper virial quantities for future work.

\begin{figure*}
\begin{center}
\includegraphics[width=\columnwidth]{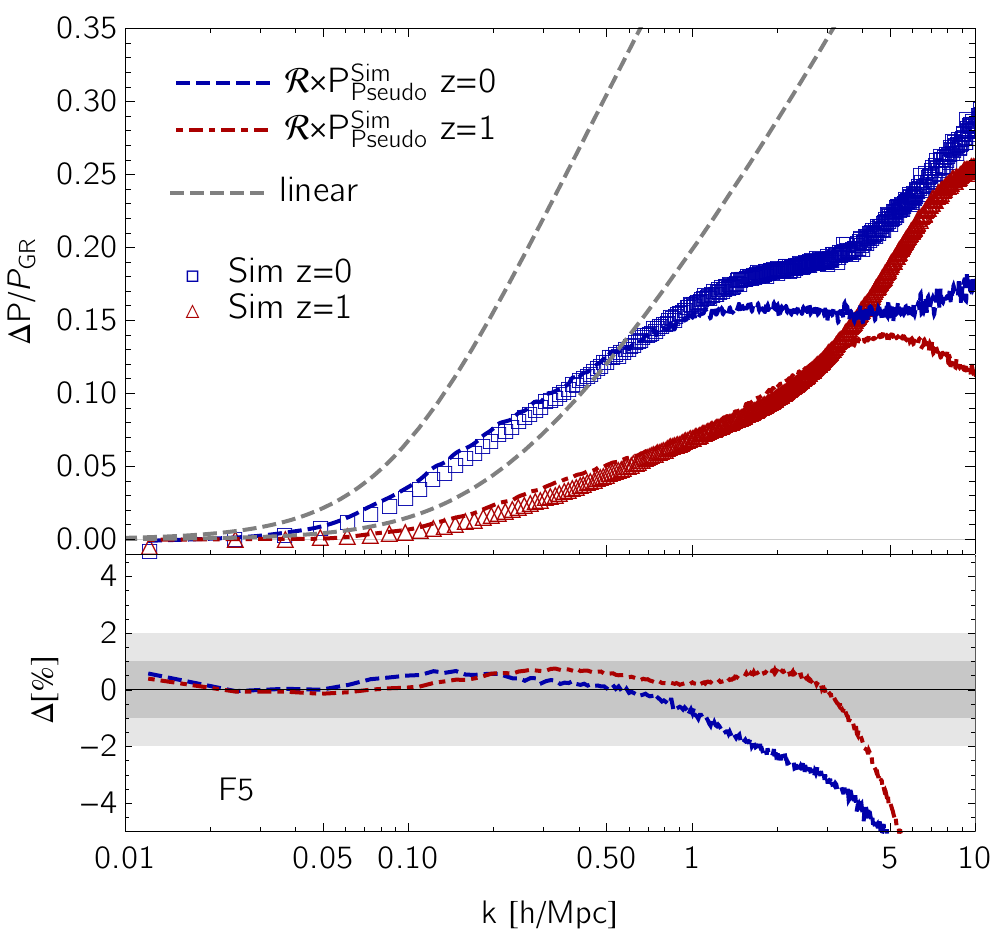}
\quad
\includegraphics[width=\columnwidth]{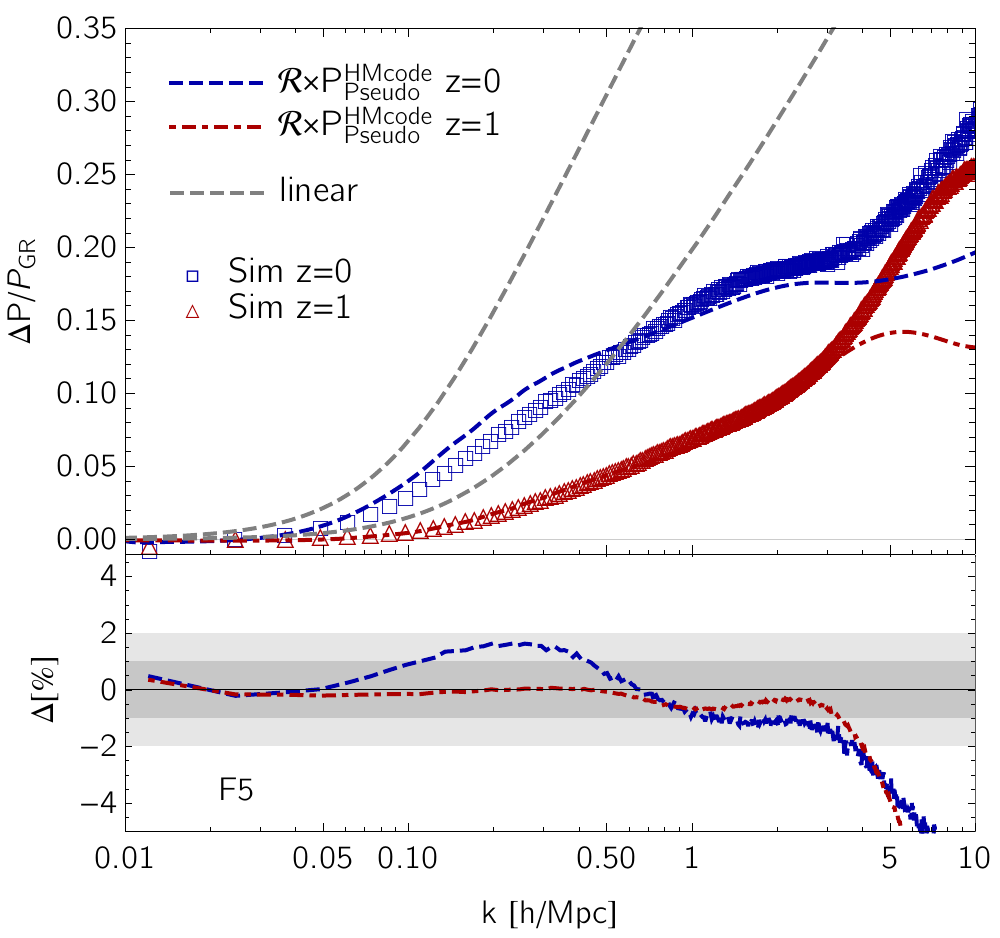}
\end{center}
\caption{Matter power spectrum fractional enhancements relative to GR for $f(R)$ gravity with $|\fRbg|=10^{-5}$. As in the previous figures, the data points correspond to the results from simulations at $z=0$ (blue squares) and $z=1$ (red triangles). Coloured lines denote predictions based on the halo model reactions at $z=0$ (dashed blue) and $z=1$ (dot-dashed red). To emphasise the impact of nonlinearities we include the linear theory predictions as dashed grey lines. Lower panels show the fractional deviation of the nonlinear predictions from the simulations, $\Delta \equiv \left(\mathcal{R} \times P_{\rm Pseudo}^{\rm Sim/HMcode}/P_{\rm GR}^{\rm Sim/HMcode} \right)/\left( P_{\rm Real}^{\rm Sim}/P_{\rm GR}^{\rm Sim} \right)-1$, with grey bands marking $1\%$ and $2\%$ uncertainty regions. {\it Left:} for our theoretical estimates we use {\it pseudo} cosmology matter power spectra measured from simulations as the baseline, which we then rescale with the halo model reactions employing the \citet{Cataneo:2016} halo mass functions. The lower panel illustrates that with future codes, eventually capable of reaching percent-level accuracy on the matter power spectra for the $\Lambda$CDM-evolved {\it pseudo} cosmologies, high-accuracy nonlinear matter power spectra in modified gravity will also be accessible. {\it Right:} same as left panel with the difference that the {\it pseudo} cosmology matter power spectra computed with \textsc{HMcode} are now adopted as the baseline. This implies that applying our halo model reaction methodology to baseline $\Lambda$CDM predictions from existing codes we can achieve $\lesssim 2\%$ precision on scales $k \lesssim 1$ \hMpc.
}
\label{fig:mpkF5}
\end{figure*}

\begin{figure*}
\begin{center}
\includegraphics[width=\columnwidth]{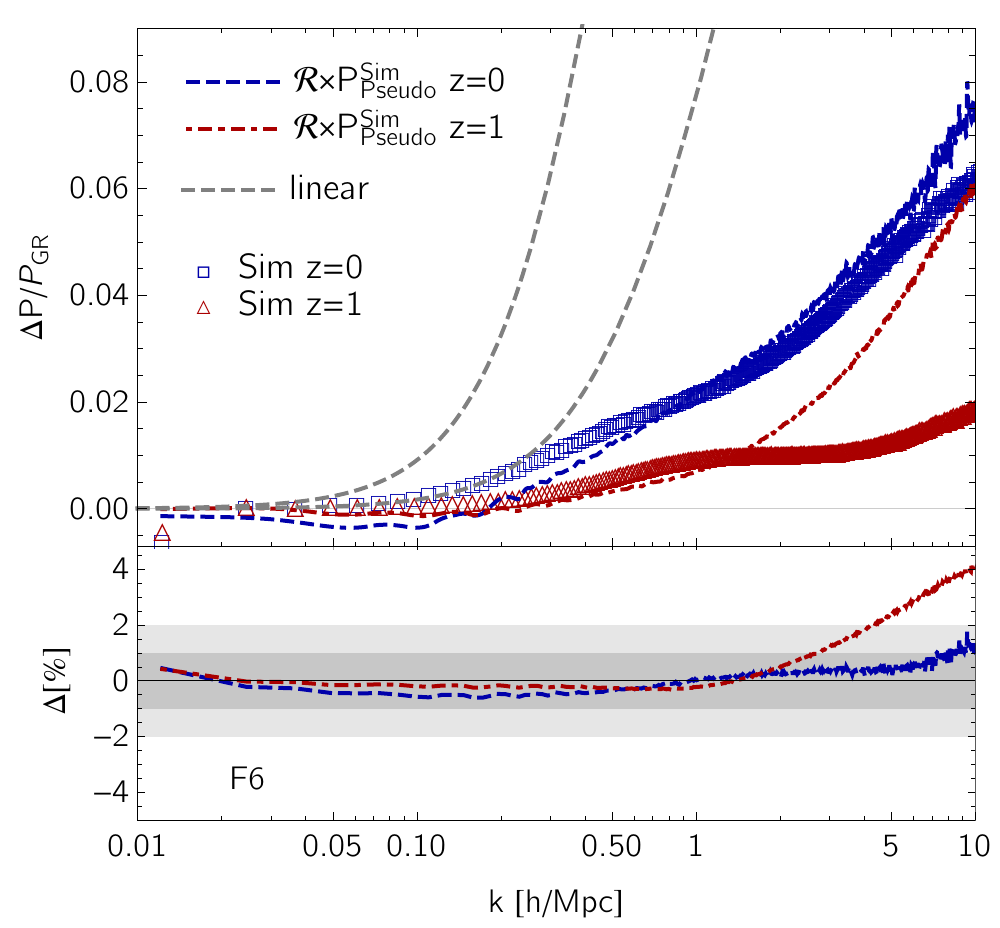}
\quad
\includegraphics[width=\columnwidth]{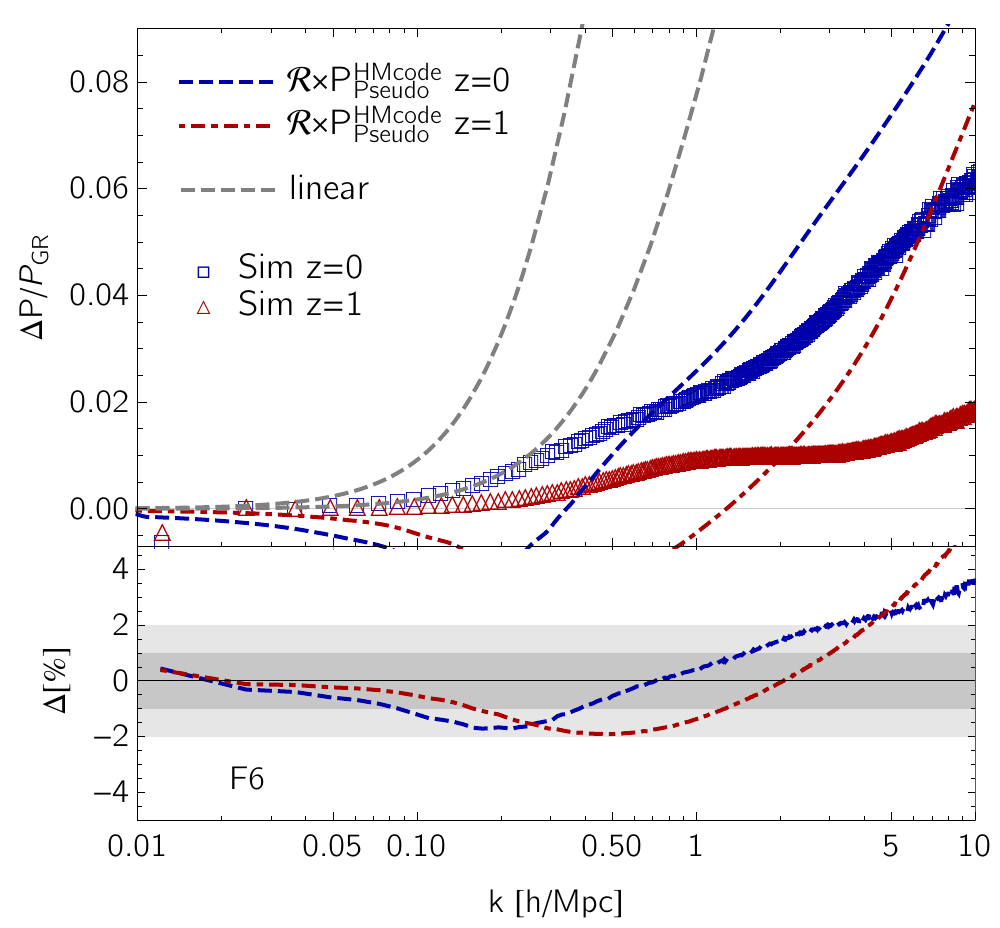}
\end{center}
\caption{Same as Figure \ref{fig:mpkF5} with the background field amplitude set to $|\fRbg|=10^{-6}$.}
\label{fig:mpkF6}
\end{figure*}

Figure~\ref{fig:mpkF5} and Figure~\ref{fig:mpkF6} show the relative change of the matter power spectrum in $f(R)$ gravity with respect to GR for the F5 and F6 models, respectively. The left panels present the best-case scenario, that is, when ``perfect'' knowledge of the \emph{pseudo} power spectrum is available. In this case, since the uncertainties come entirely from our halo model predictions, we can obviously reproduce the power spectrum ratios at the same level of accuracy of our reactions. For now the \emph{pseudo} information comes directly from our simulations, but it is not hard to imagine a specifically designed emulator capable of generating the nonlinear matter power spectrum of $\Lambda$CDM cosmologies with non-standard initial conditions. We will analyse the requirements for such emulator in a future work (Giblin et al. in prep.). In the right panels we compute $P^{\rm pseudo}$ with HM{\sc code} to demonstrate that currently, together with publicly available codes, our method can achieve 2\% accuracy on scales $k \lesssim 1$ \hMpc in modified gravity theories characterised by scale-dependent linear growth.

For a comparison to a range of other methods for modelling the nonlinear matter power spectrum in $f(R)$ and other chameleon gravity models, we refer to Figures 4 and 5 in \citet{Lombriser:2014dua}, noting that the majority of these methods rely on fitting parameters in contrast to the approach discussed here.

\subsection{DGP}

\begin{figure*}
\begin{center}
\includegraphics[width=\columnwidth]{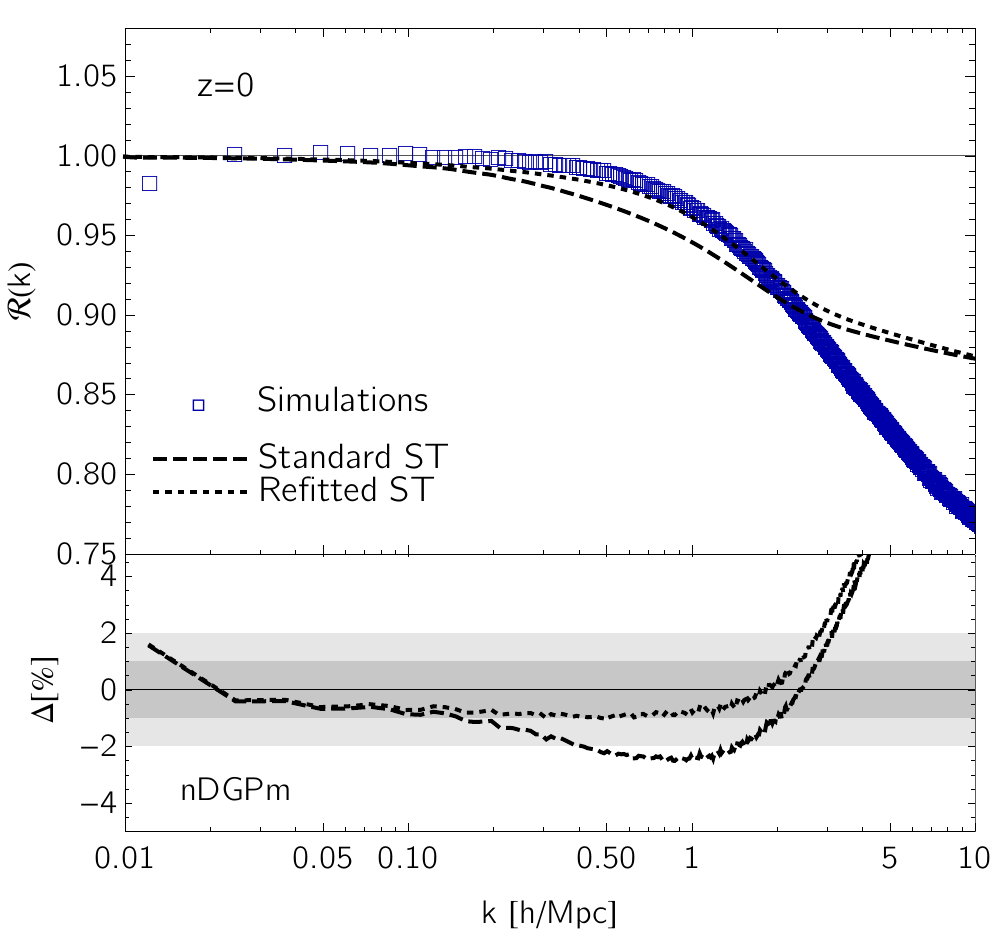}
\quad
\includegraphics[width=\columnwidth]{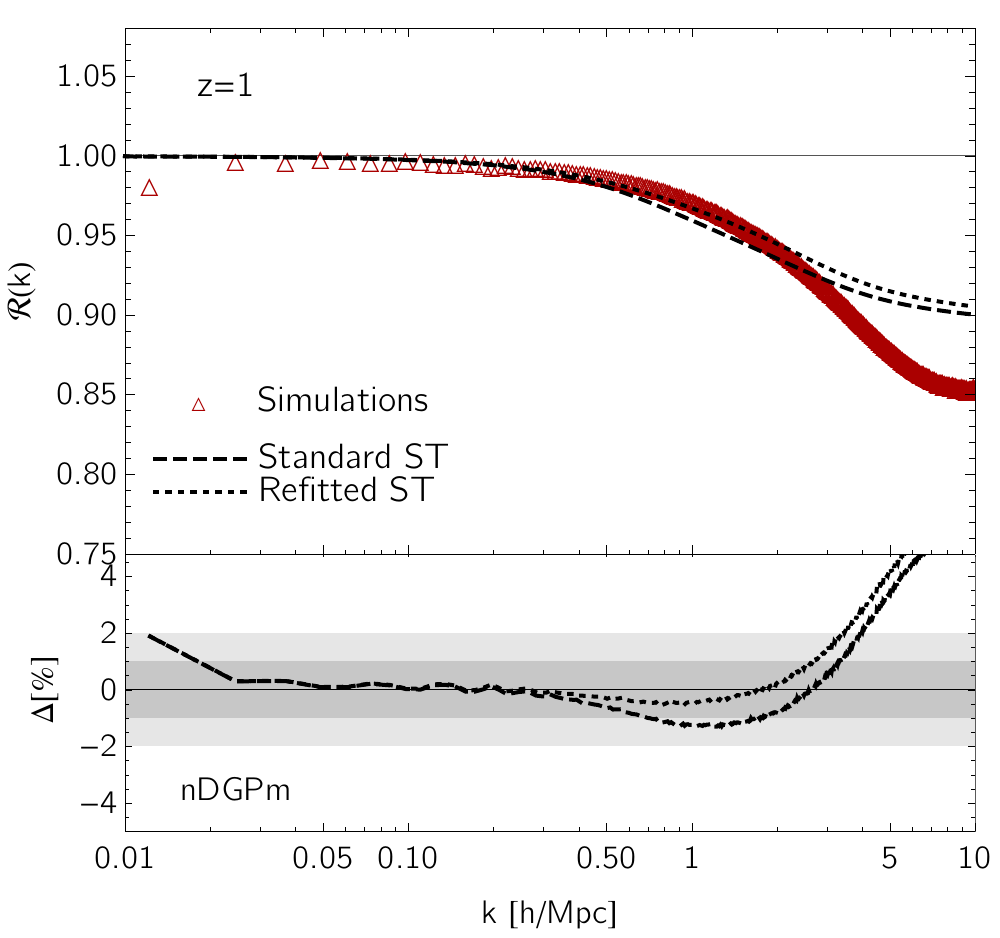}
\end{center}
\caption{Matter power spectrum reactions in an nDGP cosmology with crossover scale $r_{\rm c} H_0=0.5$. In each panel the data points represent the reactions measured from simulations; lines denote the corresponding halo model predictions defined by the halo mass function used for the one-halo contributions in nDGP gravity, that is based on either the standard Sheth-Tormen fits (dashed) or on fits to our simulations presented in App.~\ref{sec:tests} (dotted). Lower panels present the fractional deviation of the halo model relative to the simulations, with grey bands marking the $1\%$ and $2\%$ uncertainty regions. {\it Left:} reaction at $z=0$ with the refitted halo mass function significantly improving the predictions for $k \lesssim 1$ \hMpc, as shown in the lower panel.  {\it Right:} $z=1$ reaction. The lower panel shows similar performance for the two halo mass function fits, with our refitted version matching the simulations within $1\%$ over a wider range of scales.}
\label{fig:respDGPm}
\end{figure*}

\begin{figure*}
\begin{center}
\includegraphics[width=\columnwidth]{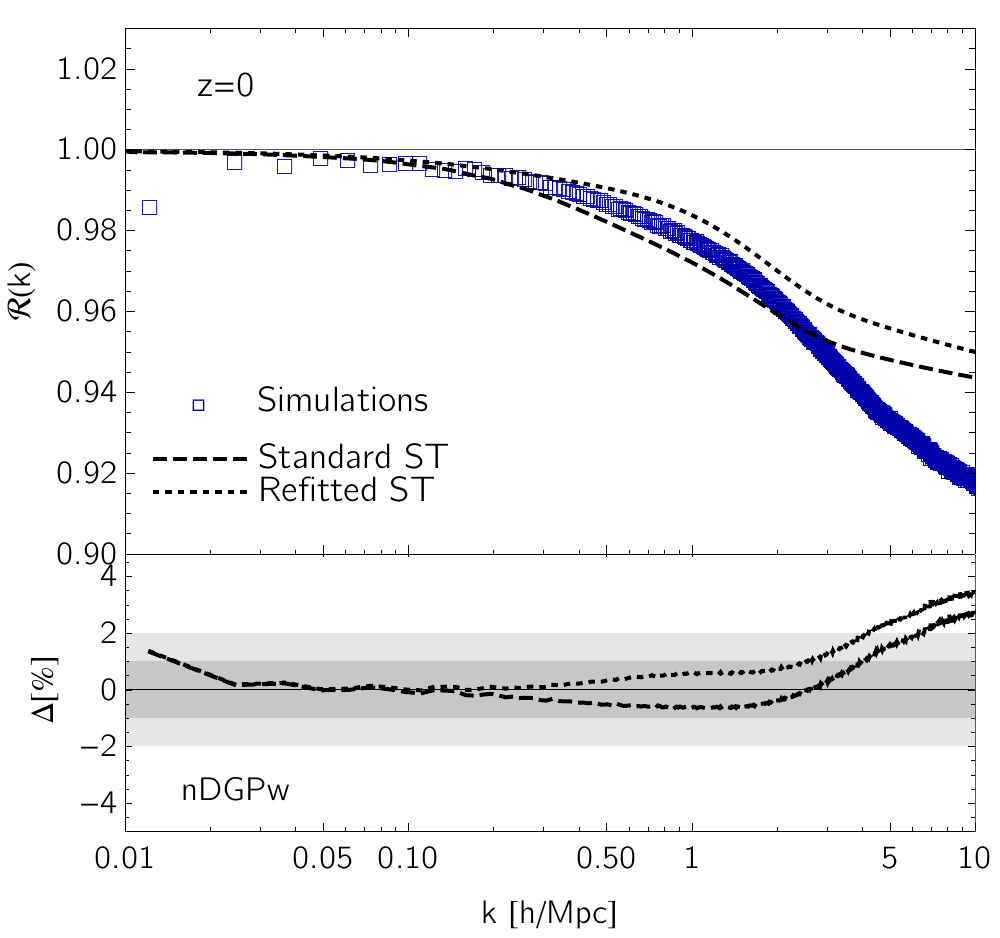}
\quad
\includegraphics[width=\columnwidth]{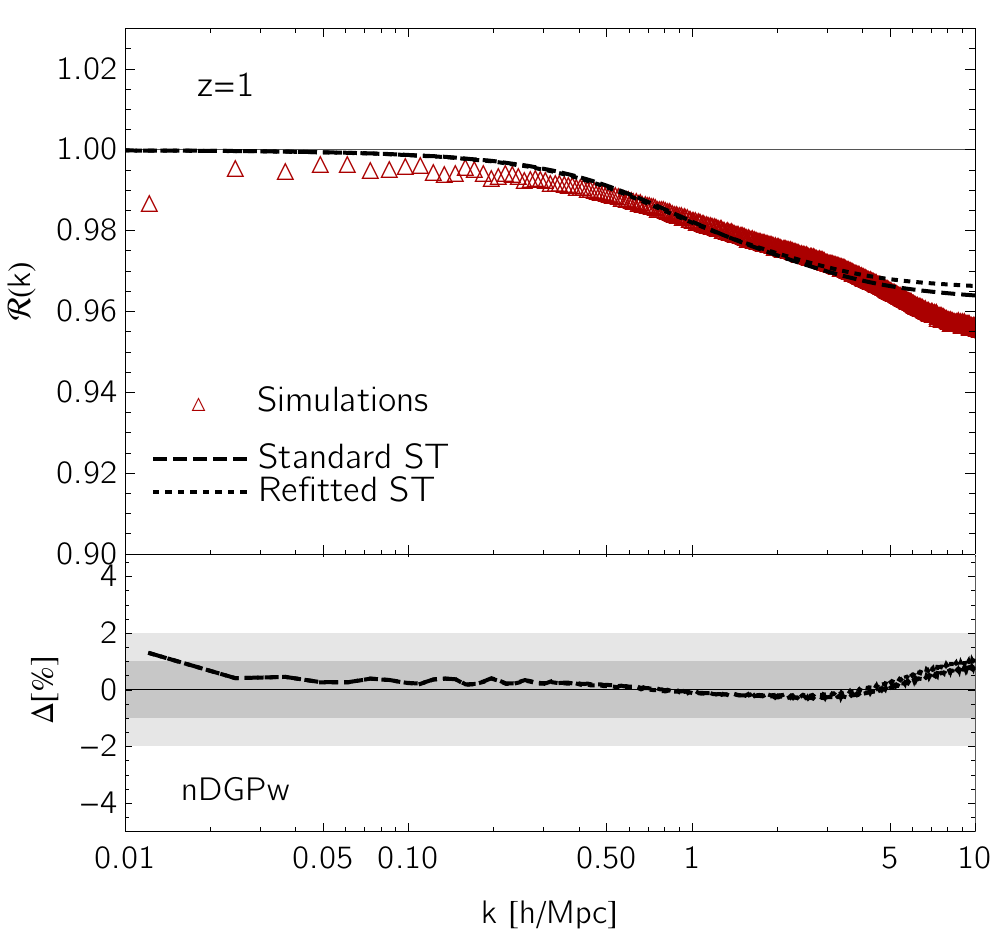}
\end{center}
\caption{Same as Figure \ref{fig:respDGPm} with the crossover scale set to $r_{\rm c} H_0=2$. Here both halo mass function fits exhibit excellent performance independent of redshift, which can be explained by the similarity of their $n_{\rm vir}^{\rm DGP}/n_{\rm vir}^{\rm Pseudo}$ ratios shown in Figure \ref{fig:P1hIgrDGPw}.}
\label{fig:respDGPw}
\end{figure*}

Spherical collapse dynamics is much simpler in nDGP~\citep{Schmidt:2010}, with both the linear overdensity threshold for collapse, $\delta_{\rm c}$, and the corresponding average virial halo overdensity, $\Delta_{\rm vir}$, being only functions of redshift. For instance, in GR one has $\Delta_{\rm vir} (z=0) = 335$ and $\Delta_{\rm vir} (z=1) = 200$ for $\Omega_{\rm m} = 0.3072$, whereas  these values become $\Delta_{\rm vir} (z=0) = 283$ and $\Delta_{\rm vir} (z=1) = 178$ in our nDGPm cosmology. This fact allows us to extract the virial halo mass function directly from our simulations, and test that accurate $n_{\rm vir}^{\rm real}/n_{\rm vir}^{\rm pseudo}$ ratios do indeed produce accurate halo model reactions. Figures~\ref{fig:respDGPm} and \ref{fig:respDGPw} show that, after refitting the virial Sheth-Tormen mass function to the same quantity from simulations, halo model predictions reach percent level accuracy on scales $k \lesssim 1$ \hMpc (see App.~\ref{sec:tests} for details on the halo mass function calibration to simulations). Moreover, since the Vainshtein radius (Eq.~\ref{eq:rv}) for the most massive halos is of order a few megaparsecs, we expect small changes caused by the screening mechanism on large scales, i.e. $k \lesssim 0.1$ \hMpc. In other words, although in our calculations we keep the two-halo correction factor in Eq.~\eqref{eq:resp_hm}, it contributes only marginally to improving the performance of our reaction functions. This is evident from the perturbation theory predictions shown in Figure~\ref{fig:SPTreacDGP}. Once again, deviations on scales $k \gtrsim 1$ \hMpc could be primarily sourced by inaccurate \emph{real} and \emph{pseudo} halo concentrations, and is the subject of future investigation.

\begin{figure*}
\begin{center}
\includegraphics[width=\columnwidth]{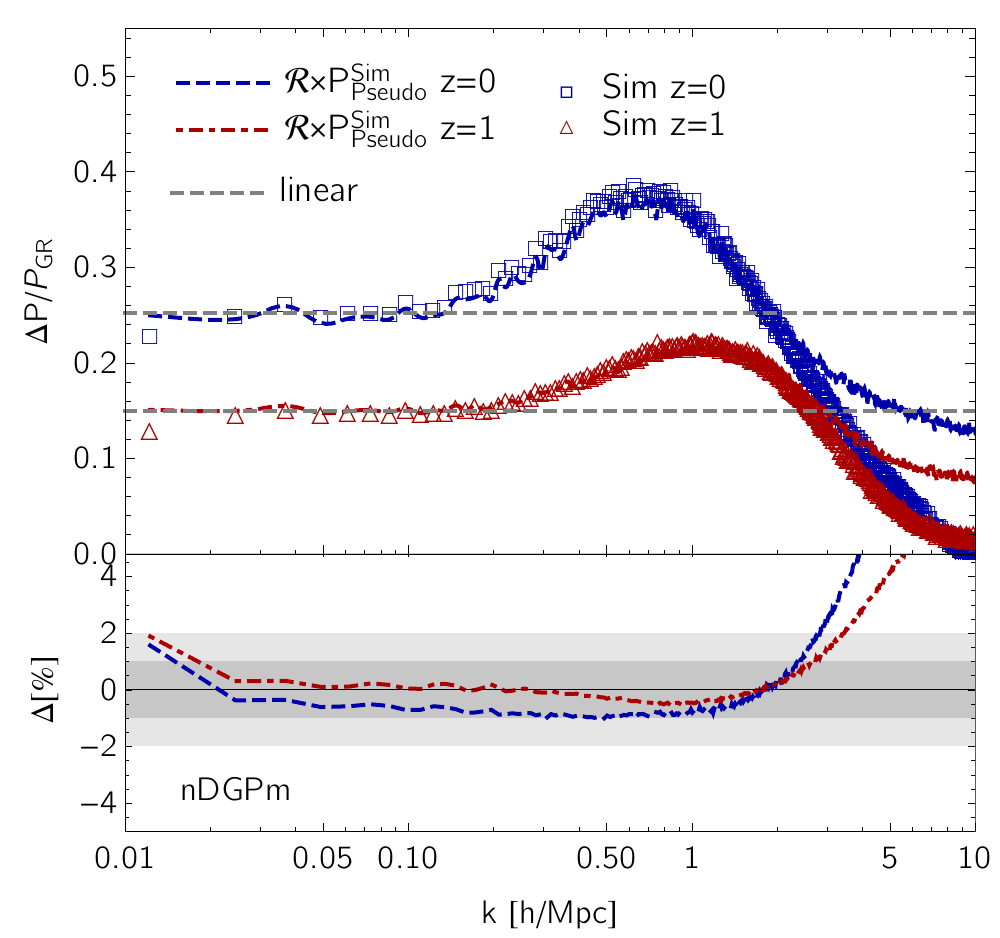}
\quad
\includegraphics[width=\columnwidth]{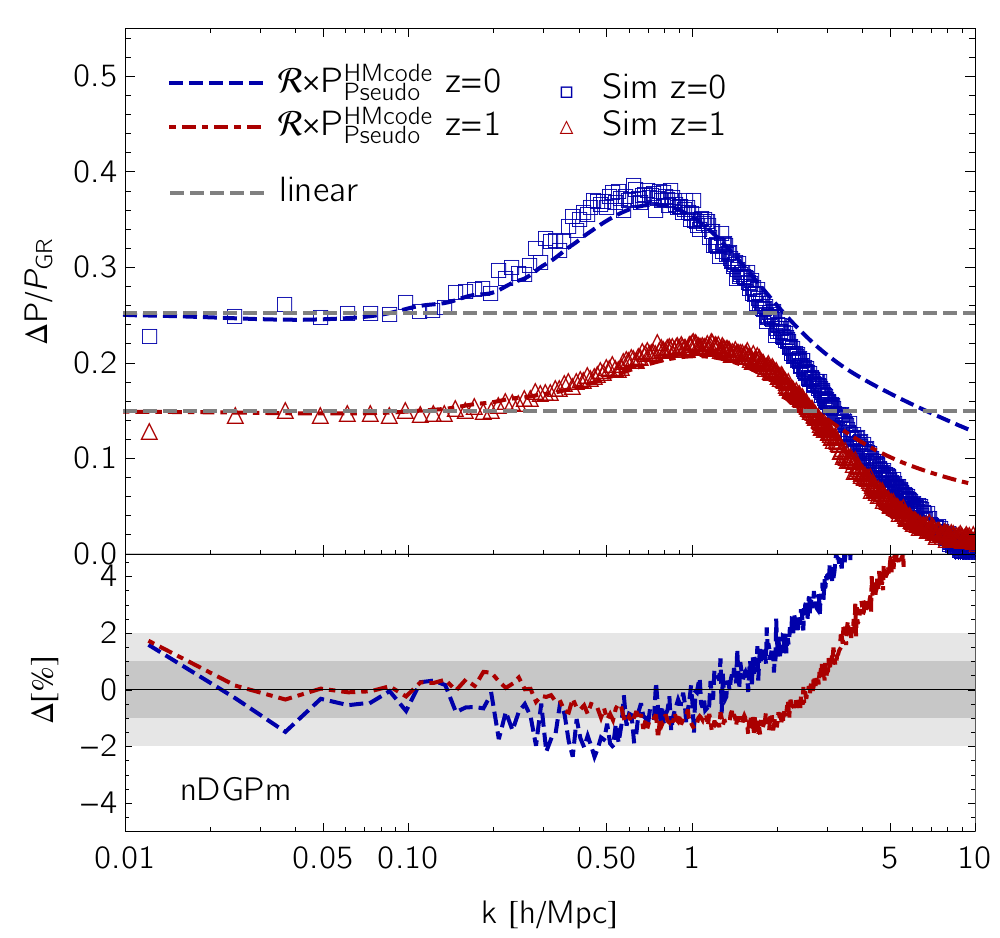}
\end{center}
\caption{Matter power spectrum fractional enhancements relative to GR for nDGP with $r_{\rm c} H_0 = 0.5$. The data points correspond to the results from simulations at $z=0$ (blue squares) and $z=1$ (red triangles). Coloured lines represent predictions based on the halo model responses at $z=0$ (dashed blue) and $z=1$ (dot-dashed red). To emphasise the impact of nonlinearities we include the linear theory predictions as dashed grey lines. Lower panels show the fractional deviation of the nonlinear predictions from the simulations, $\Delta \equiv \left(\mathcal{R} \times P_{\rm Pseudo}^{\rm Sim/HMcode}/P_{\rm GR}^{\rm Sim/HMcode} \right)/\left( P_{\rm Real}^{\rm Sim}/P_{\rm GR}^{\rm Sim} \right)-1$, with grey bands marking $1\%$ and $2\%$ uncertainty regions. {\it Left:} for our theoretical estimates we use {\it pseudo} cosmology matter power spectra measured from simulations as the baseline, which we then rescale with the halo model reactions employing our refitted halo mass functions in App.~\ref{sec:tests}. As for $f(R)$ gravity, the lower panel illustrates that with future codes eventually capable of reaching percent-level accuracy on the matter power spectra for the $\Lambda$CDM-evolved {\it pseudo} cosmologies, high-accuracy nonlinear matter power spectra for scale-independent modifications of gravity will also be within reach. {\it Right:} same as the left panel with the difference that the {\it pseudo} cosmology matter power spectra computed with \textsc{HMcode} are now adopted as the baseline. Current available codes can achieve $\lesssim 2\%$ precision on scales $k \lesssim 1$ \hMpc when used in combination with accurate halo model reactions.
}
\label{fig:mpkDGPm}
\end{figure*}

\begin{figure*}
\begin{center}
\includegraphics[width=\columnwidth]{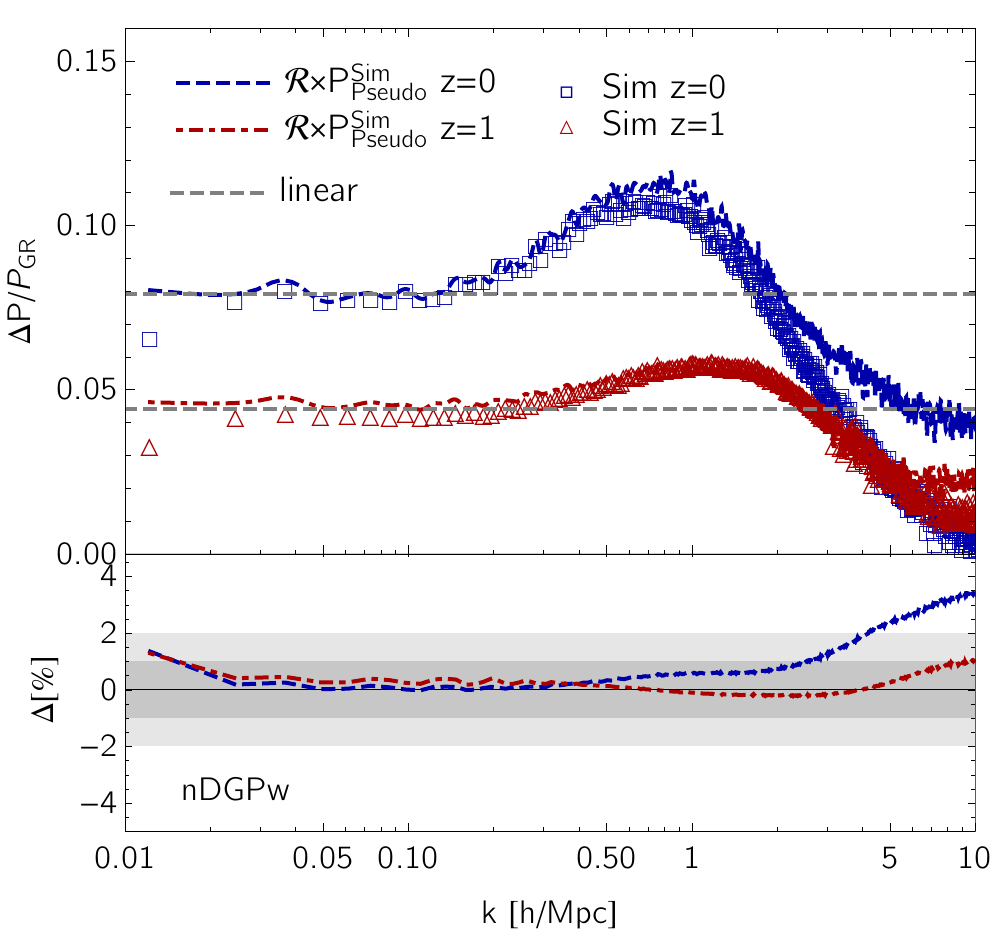}
\quad
\includegraphics[width=\columnwidth]{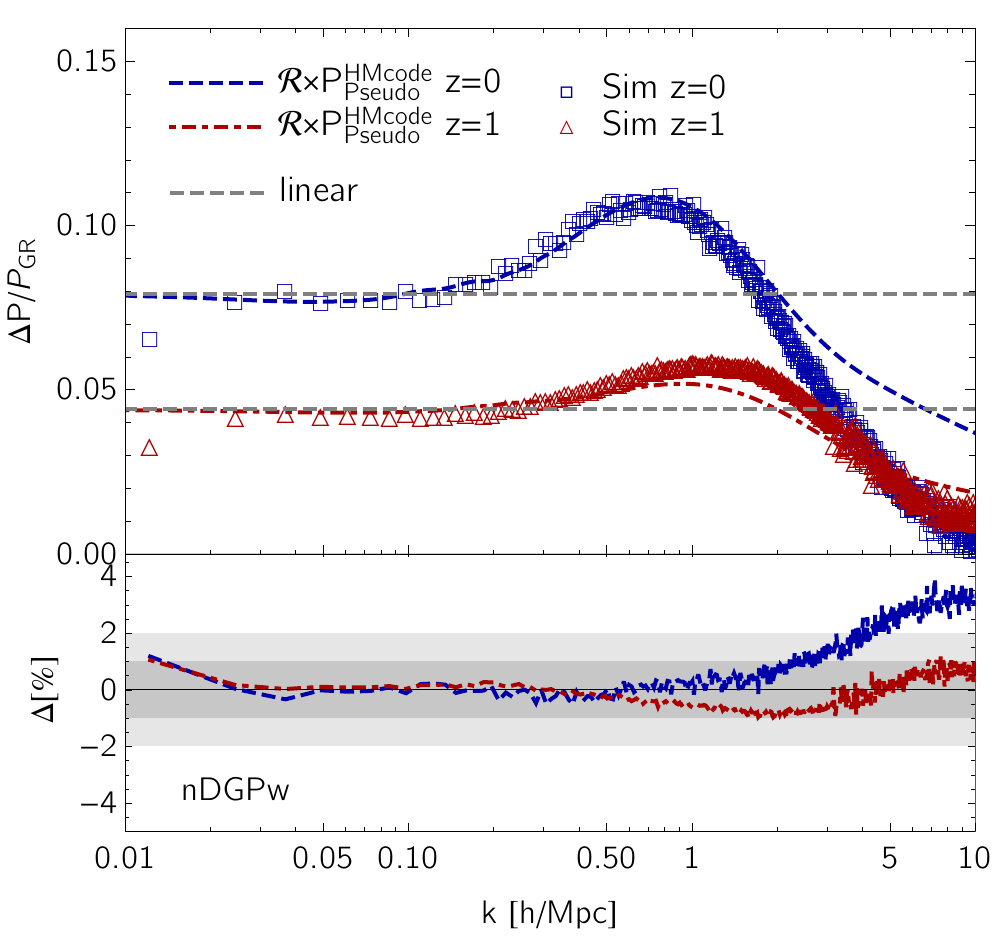}
\end{center}
\caption{Same as Figure \ref{fig:mpkDGPm} with the crossover scale set to $r_{\rm c} H_0=2$.}
\label{fig:mpkDGPw}
\end{figure*}

We study the ability of the halo model reactions to reproduce the relative difference of the nDGP power spectrum from that of standard gravity when combined with the \emph{pseudo} matter power spectrum from either HM{\sc code} or the simulations. Figures~\ref{fig:mpkDGPm} and \ref{fig:mpkDGPw} confirm that with current codes also scale-independent modifications of gravity can be predicted within 2\% over the range of scales relevant for this work.

\subsection{Dark energy}

\begin{figure*}
\begin{center}
\includegraphics[width=\columnwidth]{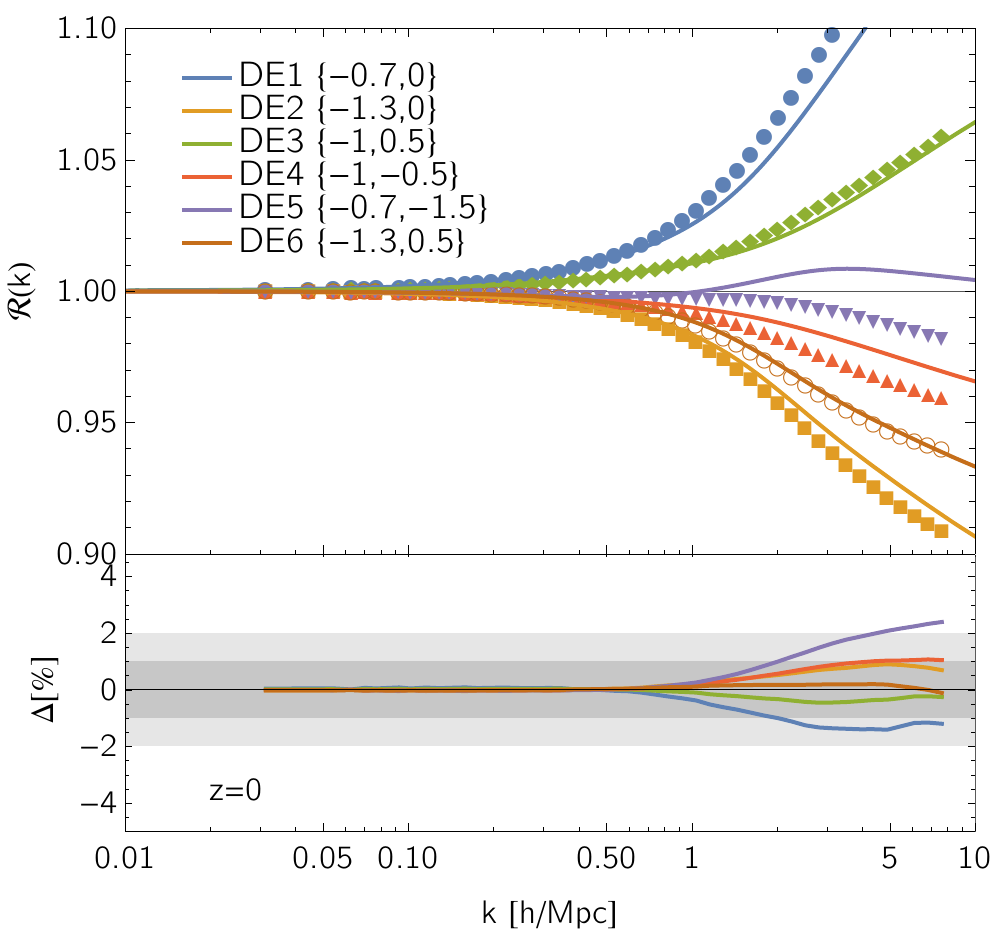}
\quad
\includegraphics[width=\columnwidth]{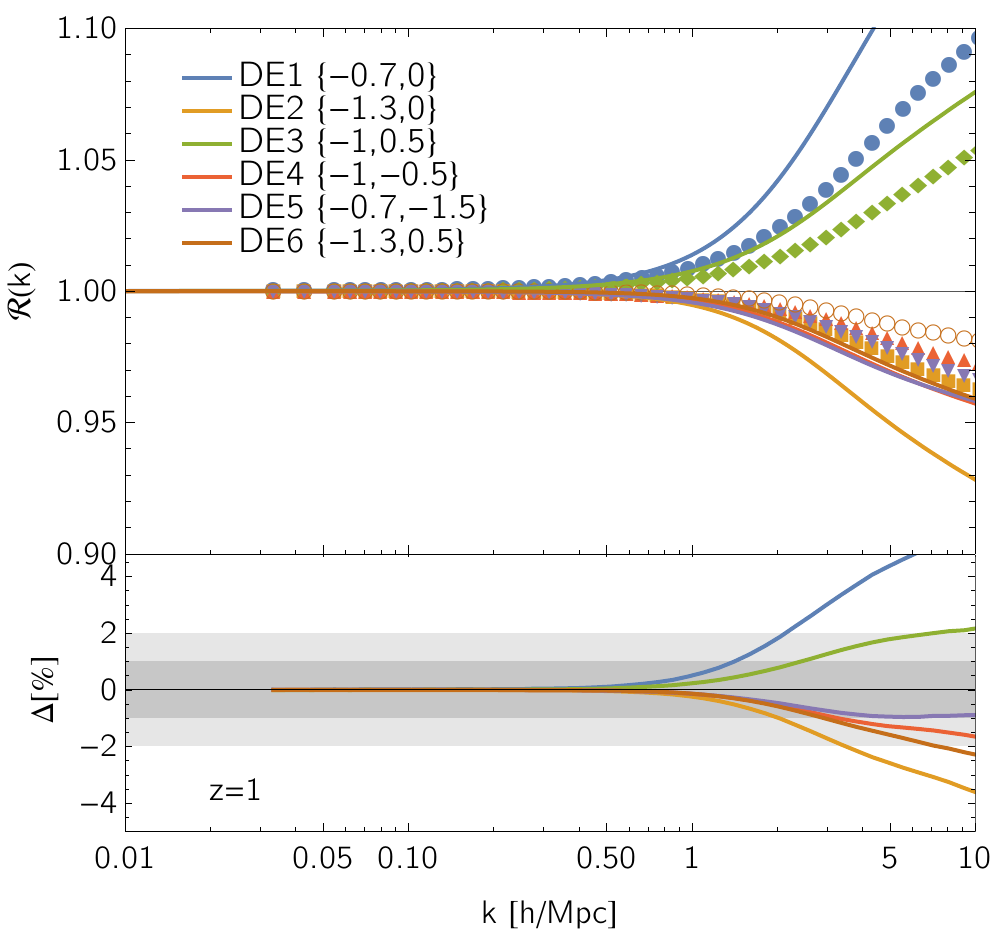}
\end{center}
\caption{Matter power spectrum reactions for the six dark energy cosmologies with $\{w_0,w_a\}$ pairs listed in Tab.~\ref{tab:de}. In each panel, the data points represent the mean reactions measured from simulations as the average from three realisations; lines denote the corresponding halo model predictions. Lower panels show the fractional deviation of the halo model relative to the simulations, with grey bands marking the $1\%$ and $2\%$ uncertainty regions. {\it Left:} reactions at $z=0$. These are similar to those presented in \citet{Mead:2017} and only differ in the derivation of the virial overdensity $\Delta_{\rm vir}$, in that here we account for all relevant contributions to the potential energy and do not assume energy conservation (see Eqs.~\ref{eq:virial_theo}-\ref{eq:KE}). On small scales agreement with the simulations is somewhat better than in modified gravity, which can be ascribed to a more accurate $c\text{-}M$ relation at $z=0$. {\it Right:} $z=1$ reactions. Although percent level accuracy is reached on scales $k \lesssim 1$ \hMpc, performance on highly non-linear scales deteriorates beyond the 2\% level for some models. We think part of the reason for that is to be found in high-$z$ inaccuracies of the \citet{Dolag:2004} prescription for the halo concentrations in dark energy cosmologies.}
\label{fig:respwCDM}
\end{figure*}

Figure~\ref{fig:respwCDM} shows the reaction functions for the evolving dark energy cosmologies listed in Table~\ref{tab:de}. The left panel contains essentially the same $z=0$ information of Figure 2 in~\cite{Mead:2017}, with the notable difference that here we compute the spherical collapse virial overdensities including the dark energy contribution to the potential energy, and do not assume energy conservation during collapse~\citep{Schmidt:2010} (expressions for the individual terms entering the virial theorem can be found in App.~\ref{sec:SC}). The right panel shows the same quantity at $z=1$. At both redshifts the halo model reactions based on the standard Sheth-Tormen mass function fits can capture very well the measurements from simulations down to the transition scale between the two- and one-halo terms. Also in this case, we attribute inaccuracies on small scales mainly to the inadequacy of the~\cite{Dolag:2004} and~\cite{Bullock:2001} halo concentration prescriptions for the \emph{real} and \emph{pseudo} cosmologies, respectively.

\begin{figure*}
\begin{center}
\includegraphics[width=\columnwidth]{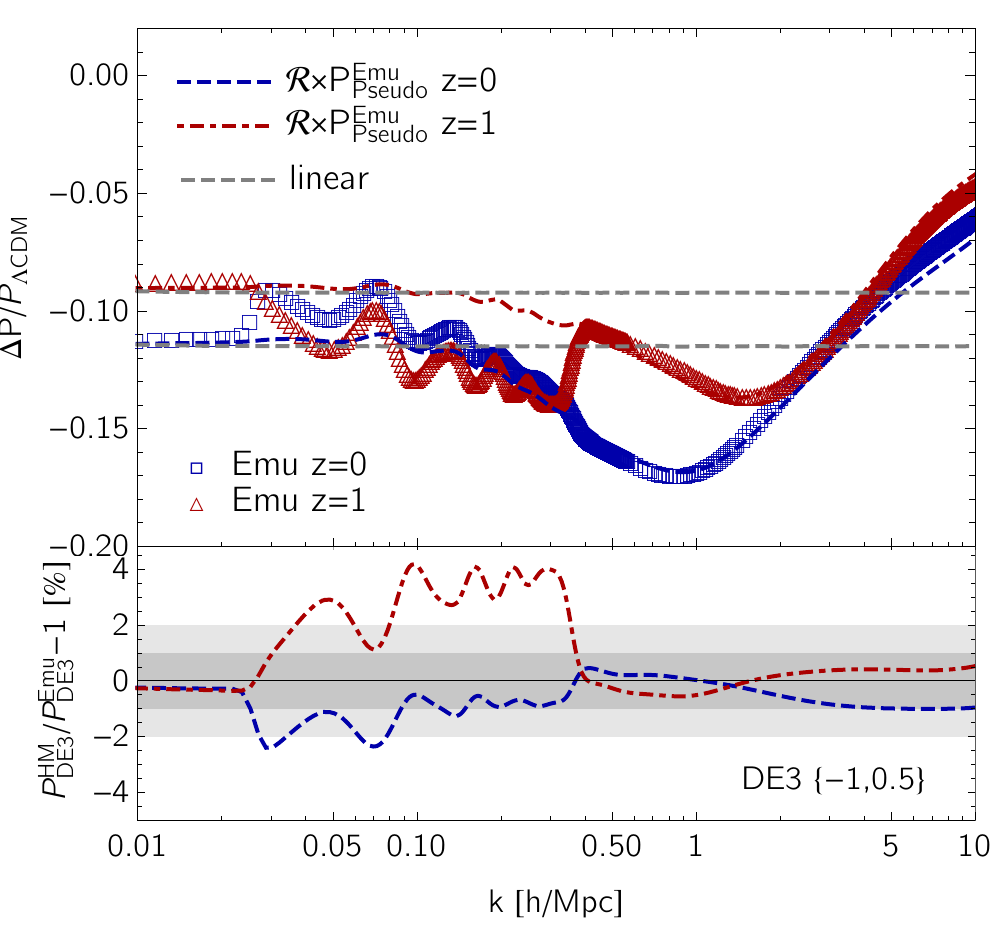}
\quad
\includegraphics[width=\columnwidth]{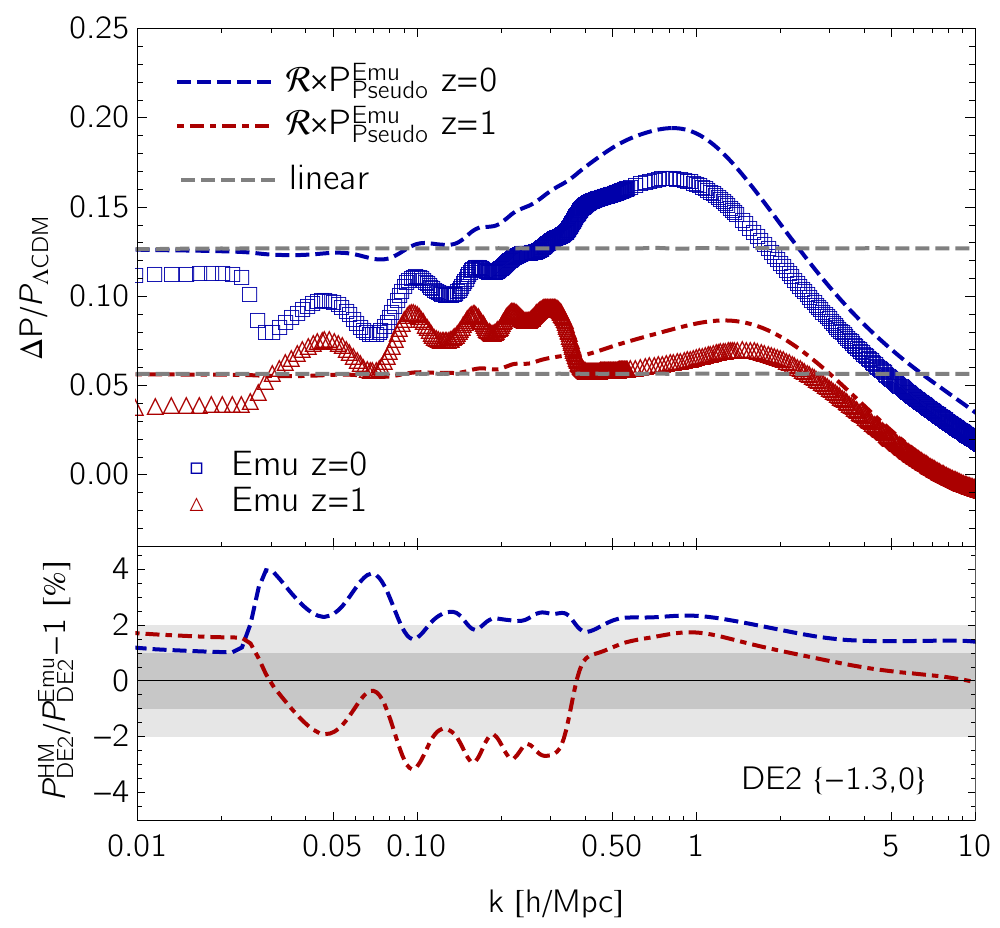}
\end{center}
\caption{Matter power spectrum fractional differences relative to $\Lambda$CDM for DE3 (left) and DE2 (right) models, where the numbers in curly brackets specify the equation of state pair $\{w_0, w_a\}$. Data points correspond to the output from the Coyote Universe emulator \citep{Heitmann:2014} at $z=0$ (blue squares) and $z=1$ (red triangles). For DE3, which has a non-constant $w$, we used the emulator extension of \citet{Casarini:2016}. Coloured lines represent predictions based on the halo model reactions at $z=0$ (dashed blue) and $z=1$ (dot-dashed red). For reference, the linear theory predictions are shown as dashed grey lines. Lower panels show the fractional deviation of the nonlinear predictions from the emulator, with grey bands marking $1\%$ and $2\%$ uncertainty regions. For our theoretical estimates we use {\it pseudo} cosmology matter power spectra computed with the emulator itself as baseline, which we then rescale with the halo model reactions. Lower panels illustrate that our halo model predictions can be employed to map accurately $\Lambda$CDM cosmologies to evolving dark energy models. Deviations of order $2\text{-}4\%$ on scales $0.02 \lesssim k \, \Mpch \lesssim 0.5$ are entirely due to the emulator being pushed to the edges of its domain of applicability in the plane $\{ \omega_b, w \}$.}
\label{fig:mpkwCDM}
\end{figure*}

In Figure~\ref{fig:mpkwCDM} we consider two representative dark energy models, DE2 and DE3, and compare their matter power spectra to that of $\Lambda$CDM with the same initial conditions. Their particular equations of state enhance the growth of structure in one case and suppress it in the other. Here, we employ the Coyote Universe Emulator~\citep{Heitmann:2014} not only for our baseline \emph{pseudo} power spectra, but also as a substitute for the \emph{real} and $\Lambda$CDM cosmology simulations. This serves as an example to illustrate a straightforward application of the reaction functions: extend the cosmological parameter space of matter power spectrum emulators designed for the concordance cosmology only, without the need to run model-dependent and expensive cosmological simulations. For the evolving equation of state of DE3 we use the emulator extension code \texttt{PKequal} built upon the work presented in~\cite{Casarini:2016}. Knowing that the output from the emulator is 1-2\% accurate on scales $k \lesssim 1$ \hMpc, from the previous results in Figure~\ref{fig:respwCDM} we can expect similar agreement between our reaction-based power spectra and those obtained from the emulator itself. This is indeed the case except in the range $0.02 \lesssim k \, \Mpch \lesssim 0.5$, where the interpolation process within the emulator fails to capture the correct dependence on $\omega_{\rm b}$ and $w$ because the specific values we use sit on the edge of its domain of applicability\footnote{The Coyote Universe emulator accepts values $0.0215 < \omega_{\rm b} < 0.0235$ and $-1.3 < w < -0.7$, while for our two evolving dark energy models we have $\omega_{\rm b} = 0.0245$, $w^{\rm (DE2)} = -1.3$ and $w_{\rm eff}^{\rm (DE3)} = -0.84$. Since our background baryon density resides outside the domain of the emulator, we set it to the maximum value allowed. This has virtually no impact on our halo model responses, in that they depend only weakly on $\omega_{\rm b}$.}.

\section{Conclusions}\label{sec:conclusions}

The spatial distribution of matter in the Universe and its evolution with time emerge from the interplay of gravitational and astrophysical processes, and are inextricably linked to the nature of the cosmic matter-energy constituents. The power spectrum is an essential statistic describing the clustering of matter in the Universe, and lies at the heart of probes of the growth of structure such as cosmic shear and galaxy clustering. Measurements of these quantities from the next generation of large-volume surveys are expected to reach percent level uncertainty -- upon careful control of systematics -- on scales where nonlinearities and baryonic physics become important. It is notoriously difficult to predict the matter power spectrum in this regime to such a degree of accuracy, yet these scales contain a wealth of information on currently unanswered questions, e.g. the nature of dark energy, the sum of neutrino masses and the extent of baryonic feedback mechanisms.
	
	In this work we focused on modelling the nonlinear matter power spectrum in modified theories of gravity and evolving dark energy cosmologies. We extended the reaction method of~\cite{Mead:2017} using the halo model to predict the nonlinear effects induced by new physics on the matter power spectrum of specifically designed reference cosmologies. These fiducial -- \emph{pseudo} -- cosmologies mimic the linear clustering of the target -- \emph{real} -- cosmologies, yet their evolution is governed by standard gravity with $\Lambda$CDM expansion histories (which are either quick to simulate with current resources, if not already available with emulators). We showed that by applying the halo model reactions to the nonlinear matter power spectrum of the \emph{pseudo} cosmologies we are able to recover the \emph{real} counterpart to within 1\% on scales $k \lesssim 1$ \hMpc for all cases under study. Remarkably, our methodology does not involve fitting the power spectra measured in simulations at any stage. Instead, having access to accurate ratios of the halo mass function in the \emph{real} cosmology to that in the \emph{pseudo} cosmology is crucial to achieve the observed performance. Not including this information from the simulation degrades the accuracy to $\lesssim 3\%$. The halo model reactions can also be used to predict the matter power spectrum in the highly nonlinear regime. However, this requires additional knowledge of the average structural properties of the dark matter halos as well as the inclusion of baryonic effects~\citep[see, e.g,][]{Schneider:2018}. We leave these improvements for future work (Cataneo et al. in prep.).
	
	In the case of the dark energy models we adopted the Coyote Universe emulator for the \emph{pseudo} matter power spectrum (i.e. for $w=-1$), which we then combined with our halo model reaction to obtain the \emph{real} expected quantity. By comparing this prediction to the \emph{real} output of the emulator (i.e. for $w\neq-1$) we showed that emulators trained on pure $\Lambda$CDM models can be accurately extended to non-standard cosmologies in an analytical way, thus substantially increasing their flexibility while simultaneously reducing the computational cost for their design. However, applications of this strategy to scale-dependent modifications of gravity (such as $f(R)$ models) necessitate of a more elaborate $\Lambda$CDM emulator that takes as input also information on the linearly modified shape of the matter power spectrum (Giblin et al. in prep.). Together with suitable halo model reactions, this emulator can also be employed to predict the nonlinear total matter power spectrum in massive neutrino cosmologies (Cataneo et al. in prep.), where the presence of a free streaming scale induces a scale-dependent linear growth~\citep{Lesgourgues:2006}.
	
	Given that our method builds on the halo model, the halo mass function and the spherical collapse model are absolutely central for our predictions. Contrary to the standard halo model calculations, however, the accuracy of our reactions strongly depends on the precision of the \emph{pseudo} and \emph{real} halo mass functions. This opens up the possibility of combining in a novel way cosmic shear and cluster abundance measurements, for example. Moreover, quite general modifications of gravity -- with their screening mechanisms -- can be implemented in the spherical collapse calculations through the nonlinear parametrised post-Friedmannian formalism of~\cite{Lombriser:2016zfz}.
	
	In summary, halo model reactions provide a fast, accurate and versatile method to compute the real-space nonlinear matter power spectrum in non-standard cosmologies. Successful implementations in redshift-space by \cite{Mead:2017} pave the way for applications to redshift-space distortions data as well. Altogether, these features make the halo model reactions an attractive alternative in-between perturbative analytical methods and brute force emulation, and promise to be an essential tool in future combined-probe analyses in search of new physics beyond the standard paradigm.

\section*{Acknowledgements}

We thank Philippe Brax, Patrick Valageas, Alejandro Aviles and Jorge Cervantes-Cota for crosschecking our perturbation theory calculations, and David Rapetti and Simon Foreman for useful discussions.
MC and CH acknowledge support from the European Research Council under grant number 647112.
LL~acknowledges support by a Swiss National Science Foundation (SNSF) Professorship grant (No.~170547), a SNSF Advanced Postdoc.Mobility Fellowship (No.~161058), the Science and Technology Facilities Council Consolidated Grant for Astronomy and Astrophysics at the University of Edinburgh, and the Affiliate programme of the Higgs Centre for Theoretical Physics.
AM has received funding from the European Union`s Horizon 2020 research and innovation programme under the Marie Sk\l{}odowska-Curie grant agreement No. 702971 and acknowledges MDM-2014-0369 of ICCUB (Unidad de Excelencia Maria de Maeztu).
SB is supported by Harvard University through the ITC Fellowship.
BL is supported by the European Research Council via grant ERC-StG-716532-PUNCA, and by UK STFC Consolidated Grants ST/P000541/1 and ST/L00075X/1.
This work used the DiRAC Data Centric system at Durham University, operated by the Institute for Computational Cosmology on behalf of the STFC DiRAC HPC Facility (\url{www.dirac.ac.uk}). This equipment was funded by BIS National E-infrastructure capital grant ST/K00042X/1, STFC capital grants ST/H008519/1, ST/K00087X/1, STFC DiRAC Operations grant ST/K003267/1 and Durham University. DiRAC is part of the National E-Infrastructure.





\bibliographystyle{mnras}
\bibliography{references} 




\appendix

\section{Spherical collapse in modified gravity and quintessence}\label{sec:SC}


We shall briefly review expressions for the force enhancement $\mathcal{F}$ in $f(R)$ and DGP gravity used in the modified spherical collapse calculation in Sec.~\ref{sec:sphericalcollapse} as well as its impact and the impact of dark energy domination on the virial theorem.

The force enhancement $\mathcal{F}$ adopted here for the spherical collapse calculation in $f(R)$ gravity is given by~\citep{Lombriser:2013}
\begin{equation}
 \mathcal{F} = \frac{1}{3} \min \left[ 3 \frac{\Delta R}{R_{\smallsub TH}} - 3 \left(\frac{\Delta R}{R_{\smallsub TH}}\right)^2 + \left(\frac{\Delta R}{R_{\smallsub TH}}\right)^3,  1 \right] \,,
\end{equation}
which uses the thin-shell approximation~\citep{Khoury:2004,Li:2012} with thickness $\Delta R\ll R_{\smallsub TH}$.
The expression is also adopted for the thick-shell limit and interpolates between the small-field ($\mathcal{F}=0$) and large-field ($\mathcal{F}=1/3$) regimes, which correspond to the two limiting scenarios studied in the $f(R)$ spherical collapse calculation of~\cite{Schmidt:2009c}.
Furthermore, for the $f(R)$ functional form Eq.~\eqref{eq:hs} one finds~\citep{Lombriser:2013}
\begin{eqnarray}
 \frac{\Delta R}{R_{\smallsub TH}} & \simeq & \frac{|f_{R0}| a^{7}}{\Omega_{\rm m} (H_0 R_{\smallsub TH})^2} y_{\rm h} \nonumber\\
 & & \times \left[ \left( \frac{1 + 4\frac{\Omega_{\Lambda}}{\Omega_{\rm m}}}{y_{\rm env}^{-3} + 4\frac{\Omega_{\Lambda}}{\Omega_{\rm m}}a^3} \right)^{2} -  \left( \frac{1 + 4\frac{\Omega_{\Lambda}}{\Omega_{\rm m}}}{y_{\rm h}^{-3} + 4\frac{\Omega_{\Lambda}}{\Omega_{\rm m}}a^3} \right)^{2} \right] \,,
 \label{eq:RRTH}
\end{eqnarray}
where the normalised top-hat radius
\be\label{eq:y_def}
y \equiv \frac{R_{\smallsub TH}/a}{R_{\rm i}/a_{\rm i}}
\ee
needs to be solved in both the halo (h) and the environment (env) using Eq.~\eqref{eq:SC1}, that now becomes  
\be\label{eq:SC2}
y^{\prime\prime}+\left(2+\frac{H^\prime}{H}\right)y^\prime  = -\frac{\Omega_{\rm m} H_0^2 a^{-3}}{2H^2}(1+\mathcal{F}) y \delta \, ,
\ee
where we set $\mathcal{F}=0$ for the $\Lambda$CDM environment and primes denote derivatives with respect to $\ln a$.
To solve Eq.~\eqref{eq:SC2} we use the initial conditions $y=1$ and $y^\prime=-\delta_{\rm i}/3$ obtained from the linear theory in an Einstein-de Sitter Universe, which at early times is an accurate description for all models in Sec.~\ref{sec:DE_MG} when the contribution from radiation is ignored. We then iteratively adjust $\delta_{\rm i}$ until the condition $y(a_{\rm coll}) = 0$ (i.e. $R_{\smallsub TH} = 0$) is satisfied at the desired time of collapse, $a_{\rm coll}$, within some small tolerance. For the environment, instead, we follow~\cite{Cataneo:2016}. Also note that Eq.~\eqref{eq:RRTH} can easily be generalised for the family of chameleon models~\citep{Lombriser:2014}. 

In DGP gravity, the force modification becomes~\citep{Schmidt:2010}
\be
 \mathcal{F} = \frac{2}{3\beta} \frac{\sqrt{1+x^{-3}}-1}{x^{-3}} \,,
\ee
where $x\equiv R_{\smallsub TH}/R_{\rm V}$, with the Vainshtein radius and $\beta$ function given in Eqs.~\eqref{eq:rv} and~\eqref{eq:beta}, respectively. In smooth quintessence cosmologies $\mathcal{F}=0$, and the sole effect on the spherical collapse dynamics enters through the non-standard background expansion. In both DGP and quintessence there is no contribution from the environment, hence $y_{\rm h}$ alone describes the full evolution. Moreover, $\mathcal{F}$ can be generalised and parametrised to cover the range of different screening mechanisms~\citep{Lombriser:2016zfz}.


The virial theorem for a general metric theory of gravity remains unchanged with respect to its formulation in GR, and reads
\be\label{eq:virial_theo}
2T+W=0 \, ,
\ee
where $W$ is the potential energy of the system and $T$ its kinetic energy.
However, energy is not conserved for evolving dark energy and modified gravity scenarios, and the virial radius cannot be related to the turnaroud radius in the usual way~\citep{Lahav:1991}. Instead, one must find the time of virialisation, $a_{\rm vir}$, that satisfies Eq.~\eqref{eq:virial_theo} when all contributions to $W$ are considered. More specifically, the Newtonian, scalar field and background potential energies take the form~\citep{Schmidt:2010}
\be\label{eq:PEs}
\frac{W_{\rm N}}{E_0} &=& -\Omega_{\rm m}\frac{a^{-1}}{a_{\rm i}^2}y^2(1+\delta) \, , \\
\frac{W_{\phi}}{E_0} &=& -\Omega_{\rm m} \frac{a^{-1}}{a_{\rm i}^2}\mathcal{F}y^2\delta \, , \\
\frac{W_{\rm eff}}{E_0} &=& -\frac{8\pi G}{3H_0^2}(1+3w_{\rm eff}) \bar\rho_{\rm eff} \frac{a^2}{a_{\rm i}^2}y^2 \, ,
\ee
where 
\be
E_0 \equiv \frac{3}{10}M(H_0R_{\rm i})^2 \, ,
\ee
and the kinetic energy can be written as
\be\label{eq:KE} 
\frac{T}{E_0} = \frac{H^2}{H_0^2} \left[ \frac{a}{a_{\rm i}} (y^\prime + y) \right]^2 \, .
\ee
Combining eqs.~\eqref{eq:virial_theo}-\eqref{eq:KE} together with Eqs.~\eqref{eq:Mvir}-\eqref{eq:vir_overdensity} allows us to find the virial radius $R_{\rm vir}$ as a function of halo mass, and for various theories of gravity as well as expansion histories.

\section{Perturbation theory}\label{sec:PT}

The nonlinear evolution of matter perturbations in modified gravity has been extensively studied in~\cite{Koyama:2009},~\cite{Brax:2012},~\cite{Brax:2013} and~\cite{Bose:2016}. Here we summarise their results and provide explicit expressions for the computation of next-to-leading-order corrections to the linear power spectrum.

The continuity and Euler equations describing the evolution of the matter density perturbations, $\delta$, and peculiar velocity, ${\bf v}$, are
\begin{gather}
\dot\delta + \frac{1}{a}\nabla \cdot [(1+\delta){\bf v}] = 0 \, , \\
\dot{\bf v} + H{\bf v} + \frac{1}{a}({\bf v} \cdot \nabla){\bf v} = -\frac{1}{a}\nabla \Psi \, ,
\end{gather}
where the gravitational potential $\Psi$ in modified gravity theories with screening mechanisms depends nonlinearly on the matter density perturbations, as in Eqs.~\eqref{eq:poisson-fr} and~\eqref{eq:gradientmod-DGP}. Assuming vanishing vorticity, the velocity field can be fully described by its divergence $\theta \equiv \nabla \cdot {\bf v}/(aH)$. By defining the vector field
\be
\varpi({\bf k},\eta) \equiv
   \begin{pmatrix} 
      \tilde\delta({\bf k},\eta) \\
      -\tilde\theta({\bf k},\eta) \\
   \end{pmatrix} \, ,
\ee
and expanding the gravitational potential up to third order in the perturbations, the fluid equations in Fourier space read
\begin{multline}\label{eq:fluid_fourier}
\varpi_i^\prime({\bf k};\eta) + \mathcal{M}_{ij}(k;\eta)\varpi_j({\bf k};\eta) = \\
	\int \frac{{\rm d}{\bf k}_1 {\rm d}{\bf k}_2}{(2\pi)^3} \delta_{\rm D}({\bf k} - {\bf k}_{12}) \gamma_{i;jk}({\bf k}_1,{\bf k}_2;\eta) \\
	\times \varpi_j({\bf k}_1;\eta)\varpi_k({\bf k}_2;\eta) \\
	+ \int \frac{{\rm d}{\bf k}_1 {\rm d}{\bf k}_2 {\rm d}{\bf k}_3}{(2\pi)^6} \delta_{\rm D}({\bf k} - {\bf k}_{123}) \sigma_{i;jkl}({\bf k}_1,{\bf k}_2;\eta) \\ 
	\times \varpi_j({\bf k}_1;\eta)\varpi_k({\bf k}_2;\eta)\varpi_l({\bf k}_3;\eta) \, ,
\end{multline}
where primes denote derivatives with respect to $\eta \equiv \ln a$, ${\bf k}_{1 \cdots n} \equiv {\bf k}_1 + \cdots +  {\bf k}_n$, and repeated indices are summed over. The left hand side of Eq.~\eqref{eq:fluid_fourier} controls the evolution of linear perturbations, with the matrix
\be
\mathcal{M}(k;\eta) = 
   \begin{pmatrix} 
      0 & -1 \\
  -\frac{3}{2}\Omega_{\rm m}(\eta)[1+\epsilon(k,\eta)] &       \frac{1-3w_{\rm eff}(\eta)\Omega_{\rm eff}(\eta)}{2} \\
   \end{pmatrix} \, .
\ee
The quantity $\epsilon$ measures the time- and scale-dependent linear departure from GR, which in $f(R)$ gravity and DGP reads
\begin{align}
\epsilon^{f(R)}(k,\eta) &= \frac{k^2}{3[e^{2\eta}m^2(\eta) + k^2]} \, , \label{eq:eps_foR}\\
\epsilon^{\rm DGP}(\eta) &= \frac{1}{3\beta(\eta)} \, , \label{eq:eps_DGP}
\end{align}
where the mass 
\be
m(\eta) = \frac{1}{\sqrt{3\kappa_1}} \, ,
\ee
with
\be
\kappa_n = \left. H^{2n-2} \frac{{\rm d}^n f_R}{{\rm d} R^n} \right|_{\bar R} \, ,
\ee
and the DGP function $\beta$ is given in Eq.~\eqref{eq:beta}.

The sources of nonlinearities are confined to the right hand side of Eq.~\eqref{eq:fluid_fourier}, where the second order vertices are 
\be
\gamma_1({\bf k}_1, {\bf k}_2; \eta) = 
	\begin{pmatrix} 
      		0 & \hat\alpha({\bf k}_2, {\bf k}_1)/2 \\
      		\hat\alpha({\bf k}_1, {\bf k}_2)/2 & 0 \\
   	\end{pmatrix} \, ,
\ee
\be
\gamma_2({\bf k}_1, {\bf k}_2; \eta) = 
	\begin{pmatrix} 
      		\gamma_{2;11}({\bf k}_1, {\bf k}_2;\eta) & 0 \\
      		0 & \hat\beta({\bf k}_1, {\bf k}_2) \\
   	\end{pmatrix} \,
\ee
with 
\begin{align}
\hat\alpha({\bf k}_1, {\bf k}_2) &= \frac{({\bf k}_1 + {\bf k}_2) \cdot {\bf k}_1}{k_1^2} \, , \\
\hat\beta({\bf k}_1, {\bf k}_2) &= \frac{\|{\bf k}_1 + {\bf k}_2\|^2 ({\bf k}_1 \cdot {\bf k}_2)}{2k_1^2 k_2^2} \, ,
\end{align}
and the only modified gravity non-vanishing third order vertex is $\sigma_{2;111}$, with
\begin{multline}
\sigma_{2;111}^{f(R)}({\bf k}_1, {\bf k}_2, {\bf k}_3;\eta) =  \\ 
 	\frac{9e^{6\eta}m^8\Omega_{\rm m}^3 k_{123}^2}{4(e^{2\eta}m^2+k_1^2)(e^{2\eta}m^2+k_2^2)} \\
	\times \frac{e^{2\eta}m^2\kappa_3 + (\kappa_3 - 9m^2\kappa_2^2)\|{\bf k}_2 + {\bf k}_3\|^2}{(e^{2\eta}m^2 + k_3^2)(e^{2\eta}m^2 + \|{\bf k}_2 + {\bf k}_3\|^2)(e^{2\eta}m^2 + k_{123}^2)} \, ,
\end{multline}
\be
\sigma_{2;111}^{\rm DGP}({\bf k}_1, {\bf k}_2, {\bf k}_3;\eta) = 0 \, .
\ee
The vertex $\gamma_{2;11}=0$ in standard gravity, whereas
\begin{multline}
\gamma_{2;11}^{f(R)}({\bf k}_1, {\bf k}_2;\eta) = \\ 
	\frac{9e^{4\eta}\Omega_{\rm m}^2m^6\kappa_2 k_{12}^2}{4(e^{2\eta} m^2 + k_{12}^2)(e^{2\eta} m^2 + k_1^2)(e^{2\eta} m^2 + k_2^2)} \, ,
\end{multline}
\be
\gamma_{2;11}^{\rm DGP}({\bf k}_1, {\bf k}_2;\eta) = \frac{1}{6}\Omega_{\rm m}^2 \frac{H^2}{H_0^2}\frac{(H_0r_{\rm c})^2}{\beta^2} \left[ \frac{({\bf k}_1 \cdot {\bf k}_2)^2}{k_1^2 k_2^2} -1 \right] \, .
\ee

The one-loop power spectrum corrections require the density and velocity fields up to third order in the perturbations, that is, we need $\varpi = \varpi^{(1)} + \varpi^{(2)} + \varpi^{(3)}$. For this, we shall first find the linear solution to Eq.~\eqref{eq:fluid_fourier}, $\varpi^{(1)}$, and then derive recursively the higher order solutions with the help of the retarded Green function, which is obtained by setting the right hand side of Eq.~\eqref{eq:fluid_fourier} to a Dirac delta function~\cite[see][for more details]{Brax:2012,Brax:2013}. We can write the linear ansatz as
\be
\varpi^{(1)}({\bf k},\eta) = \tilde\delta({\bf k},0)
   \begin{pmatrix} 
      D_{+}(k,\eta) \\
      D_{+}^\prime(k,\eta) \\
   \end{pmatrix} \, ,
\ee
where the growing mode, $D_{+}(k,\eta)$, satisfies the linear second-order equation
\begin{align}
D_{+}^{\prime\prime} + \left( \frac{1-3w_{\rm eff}(\eta)\Omega_{\rm eff}(\eta)}{2} \right) D_{+}^{\prime} - \frac{3}{2}\Omega_{\rm m}(\eta) [1+ \epsilon(k,\eta)]D_{+} = 0 \, ,
\end{align}
with Einstein-de Sitter initial conditions $D_{+}(\eta_{\rm i}) = D_{+}^{\prime}(\eta_{\rm i}) = e^{\eta_{\rm i}}$ at the initial time $\eta_{\rm i}$. The decaying mode, $D_{-}$, also satisfies this equations and can be directly computed as
\be
D_{-}(k,\eta) = -D_{+}(k,\eta) \int_\eta^\infty {\rm d} \eta^\prime \frac{W(\eta^\prime)}{D_{+}(k,\eta^\prime)^2} \, ,
\ee
where the Wronskian of $D_{+}$ and $D_{-}$ is given by
\be
W(\eta) = -e^{-(1/2)\int_0^\eta {\rm d} \eta^\prime [1-3w_{\rm eff}(\eta^\prime)\Omega_{\rm eff}(\eta^\prime)]} \, .
\ee
By using the retarded Green function
\begin{multline}\label{eq:green_func}
\mathcal{G}(k_1,k_2;\eta_1,\eta_2) = \Theta(\eta_1-\eta_2) \\
	\times (2\pi)^3 \delta_{\rm D}({\bf k}_1 - {\bf k}_2) \mathcal{\tilde G}(k_1,k_2;\eta_1,\eta_2) \, ,
\end{multline}
where $\Theta(\eta_1-\eta_2)$ denotes the Heaviside function, we obtain the second and third order density perturbations as the convolutions
\be
\tilde\delta^{(2)}({\bf k},\eta) &= \mathcal{G}_{1m} \ast [\mathcal{K}^{(2)}_{m,ij} \ast \varpi_i^{(1)} \varpi_j^{(1)}],
\ee
\begin{multline}
\tilde\delta^{(3)}({\bf k},\eta) = 2\mathcal{G}_{1m} \ast [\mathcal{K}^{(2)}_{m,ij} \ast \varpi_i^{(2)} \varpi_j^{(1)}] \\
 + \mathcal{G}_{1m} \ast [\mathcal{K}^{(3)}_{m,ijk} \ast \varpi_i^{(1)} \varpi_j^{(1)} \varpi_k^{(1)}] \, ,
\end{multline}
with the operators $\mathcal{K}^{(2)}$ and $\mathcal{K}^{(3)}$ built from the second and third order vertices, $\gamma_m$ and $\sigma_m$, respectively~\cite[see][for more details]{Brax:2013}. In Eq.~\eqref{eq:green_func} the matrix $\mathcal{\tilde G}$ takes the explicit form
\begin{multline}
\mathcal{\tilde G}(k_1,k_2;\eta_1,\eta_2) = \frac{1}{D_{+2}^\prime D_{-2} - D_{-2}^\prime D_{+2}} \\ 
   \times
   \begin{pmatrix} 
      D_{+1}^\prime D_{-2} - D_{-2}^\prime D_{+1} & D_{-2}D_{+1} - D_{+2}D_{-1} \\
      D_{+2}^\prime D_{-1}^\prime - D_{-2}^\prime D_{+1}^\prime & D_{-2} D_{+1}^\prime - D_{+2} D_{-1}^\prime \\
   \end{pmatrix} \, ,
\end{multline}
where $D_{+/-i}$ is shorthand for $D_{+/-}(k_i,\eta_i)$.

Finally, we can write the equal-time two-point correlation -- equivalent to Eq.~\eqref{eq:P1loop} --  as
\be
\langle \tilde\delta \tilde\delta \rangle = \langle \tilde\delta^{(1)} \tilde\delta^{(1)} \rangle + \langle \tilde\delta^{(2)} \tilde\delta^{(2)} \rangle + \langle \tilde\delta^{(1)} \tilde\delta^{(3)} \rangle + \langle \tilde\delta^{(3)} \tilde\delta^{(1)} \rangle \, ,
\ee
which produces the following expressions for the one-loop integrals
\be
P_{22}(k,\eta) &=& 2 \int \frac{{\rm d}\bf q}{(2\pi)^3} \int_{-\infty}^\eta {\rm d}\eta^\prime \int_{-\infty}^{\eta} {\rm d}\eta^{\prime\prime} \nonumber \\
		 	&&\times \, \mathcal{\tilde G}_{1j}(k,k;\eta,\eta^\prime) \mathcal{\tilde G}_{1i}(q,q;\eta,\eta^{\prime\prime}) \nonumber \\
			&&\times \, \tilde C_{j_1j_2}(q,\| {\bf k - q} \|; \eta^\prime,\eta^\prime) \tilde C_{i_1i_2}(q,\| {\bf k - q} \|; \eta^{\prime\prime},\eta^{\prime\prime}) \nonumber \\
			&&\times \, \gamma_{j;j_1j_2}({\bf q},{\bf k - q};\eta^\prime) \gamma_{i;i_1i_2}({\bf q},{\bf k - q};\eta^{\prime\prime}) \nonumber \\
			&&\times \, P_{\smallsub L}(q) P_{\smallsub L}(\| {\bf k - q} \|) \, ,
\ee
\be
P_{13}(k,\eta) &=& 8P_{\smallsub L}(k) \int \frac{{\rm d}\bf q}{(2\pi)^3} \int_{-\infty}^\eta {\rm d}\eta^\prime \int_{-\infty}^{\eta^\prime} {\rm d}\eta^{\prime\prime} \nonumber \\
		 	&&\times \, \mathcal{\tilde G}_{1i_1}(k,k;\eta,\eta^\prime) \mathcal{\tilde G}_{j_1j_2}(q,q;\eta^\prime,\eta^{\prime\prime}) \nonumber \\
			&&\times \, \tilde C_{m_1m_2}(\| {\bf k - q} \|,\| {\bf k - q} \|; \eta^\prime,\eta^{\prime\prime}) \nonumber \\
			&&\times \, \tilde C_{1j_2}(k,k; \eta,\eta^{\prime\prime}) \gamma_{i_1;j_1m_1}({\bf q},{\bf k - q};\eta^\prime) \nonumber \\
			&&\times \, \gamma_{i_2;j_2m_2}({\bf k},{\bf q - k};\eta^{\prime\prime}) P_{\smallsub L}(\| {\bf k - q} \|) \, ,
\ee
\be
P_{13}^{\Psi}(k,\eta) &=& 6P_{\smallsub L}(k) \int \frac{{\rm d}\bf q}{(2\pi)^3} \int_{-\infty}^\eta {\rm d}\eta^\prime \mathcal{\tilde G}_{12}(k,k;\eta,\eta^\prime) \nonumber \\
				&&\times \, \tilde C_{11}(k,k;\eta,\eta^\prime) \tilde C_{11}(q,q;\eta^\prime,\eta^\prime) \nonumber \\
				&&\times \, \sigma_{2;111} ({\bf q , -q, k};\eta^\prime) P_{\smallsub L}(q) \, ,
\ee
where all indices run from 1 to 2, the linear power spectra are taken at $\eta=0$, and we have used the two-point linear correlator
\be
   \tilde C(k_1,k_2;\eta_1,\eta_2) =
   \begin{pmatrix} 
      D_{+1}D_{+2} & D_{+1}D_{+2}^\prime \\
      D_{+1}^\prime D_{+2} & D_{+1}^\prime D_{+2}^\prime \\
   \end{pmatrix} \, .
\ee
\begin{figure*}
\begin{center}
\includegraphics[width=\columnwidth]{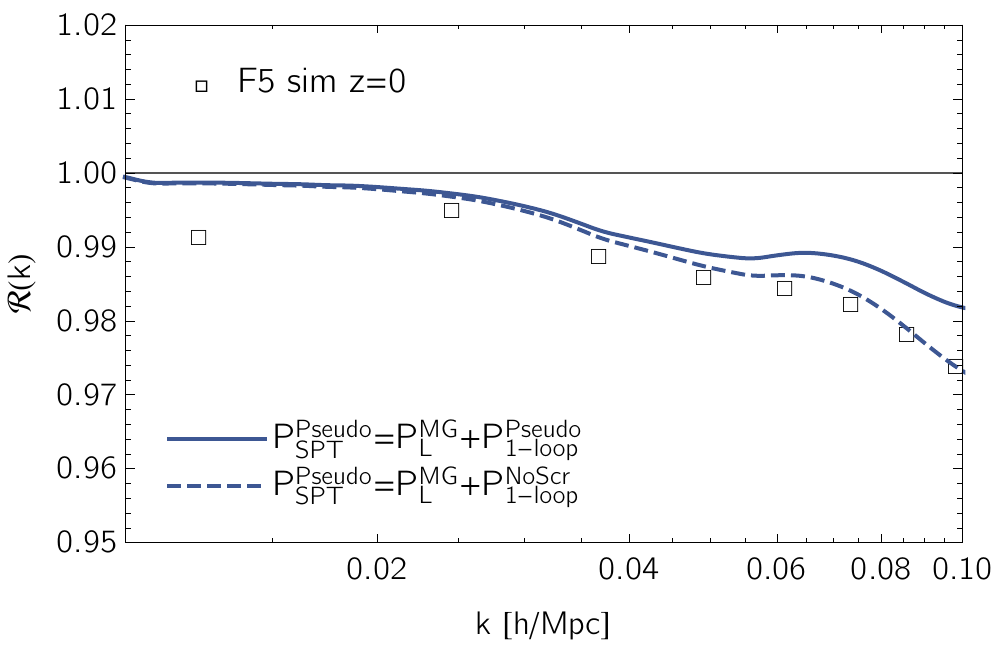}
\quad
\includegraphics[width=\columnwidth]{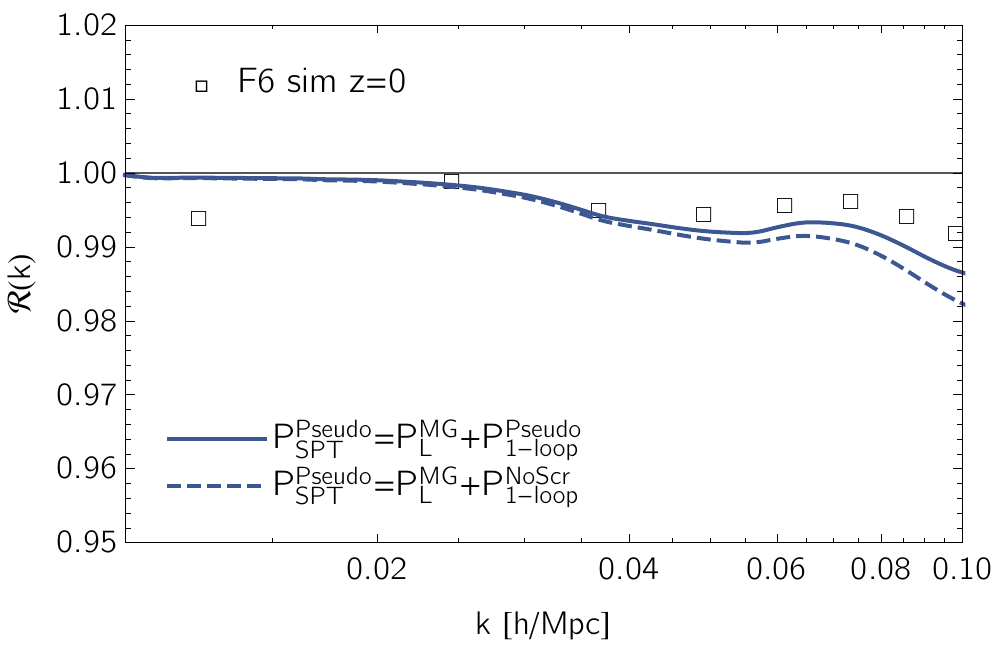}
\end{center}
\caption{Standard perturbation theory (SPT) matter power spectrum reactions in $f(R)$ gravity with background scalaron amplitudes $|\bar f_{R0}|=10^{-5}$ (left) and $|\bar f_{R0}|=10^{-6}$ (right). The different lines illustrate the effect of computing the nonlinear power spectrum for the {\it pseudo} cosmology using either the exact one-loop corrections (solid) or the {\it real} unscreened one-loop terms instead (dashed). The squares are the measurements from our simulations. In F5 the {\it real} screened and unscreened one-loop contributions err in the same direction, which makes the no-chameleon correction preferable over the theoretically motivated {\it pseudo} one-loop term. The opposite is true for F6, although the difference between the two approaches is small in this case. For consistency, throughout this work we always choose the `no-screening' one-loop correction for our {\it pseudo} SPT predictions, a strategy good enough on large quasi-linear scales for all cosmologies investigated.}
\label{fig:SPTreacFoR}
\end{figure*}
\begin{figure*}
\begin{center}
\includegraphics[width=\columnwidth]{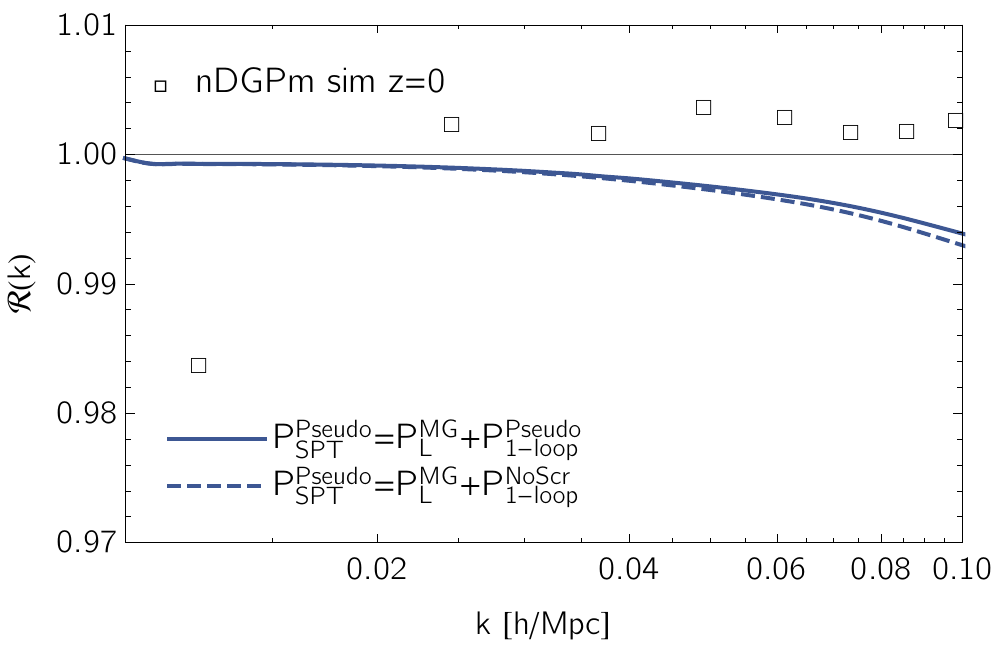}
\quad
\includegraphics[width=\columnwidth]{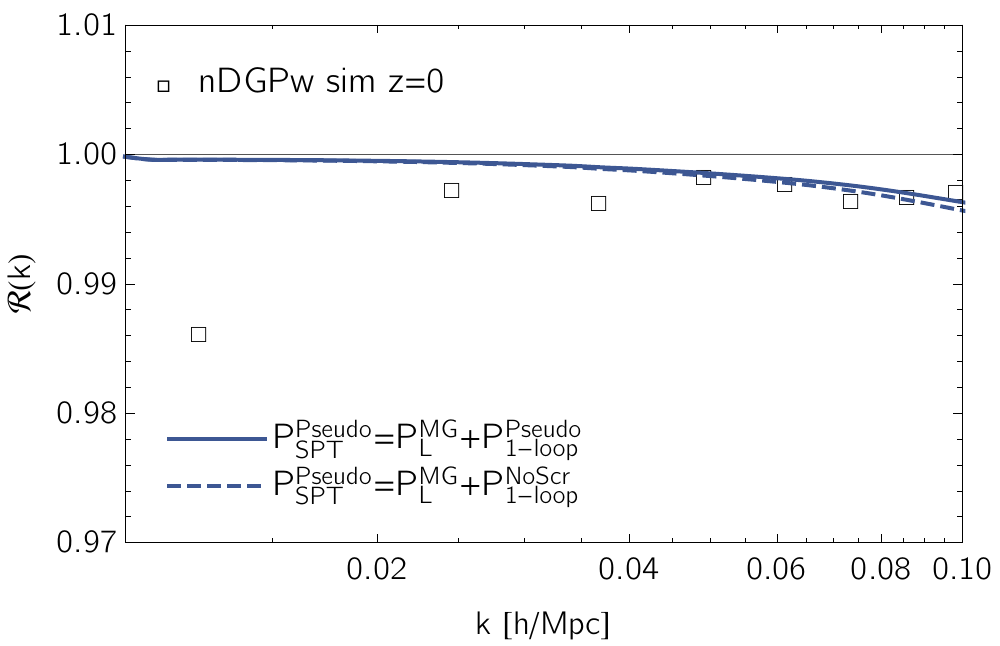}
\end{center}
\caption{Same as Figure \ref{fig:SPTreacFoR} for nDGP gravity with crossover scales $r_{\rm c} H_0=0.5$ (left) and $r_{\rm c} H_0=2$ (right). Here using the {\it real} unscreened one-loop corrections instead of the equivalent exact {\it pseudo}  quantity has negligible impact on the SPT responses.}
\label{fig:SPTreacDGP}
\end{figure*}
To deactivate the screening mechanisms we simply set $\gamma_{2;11} = \sigma_{2;111} = 0$, while keeping the linear deviation $\epsilon$ given in Eqs.~\eqref{eq:eps_foR}-\eqref{eq:eps_DGP}. This results in what we call the \emph{real} `no-screening' power spectrum, $P_{\rm NoScr}^{\rm real}$. The \emph{pseudo} power spectrum up to one-loop, instead, follows from also imposing $\epsilon = 0$, such that the only difference compared to the standard cosmology is in the initial conditions, i.e. in the shape and/or amplitude of the linear power spectrum. Figures~\ref{fig:SPTreacFoR} and~\ref{fig:SPTreacDGP} show the SPT reaction functions on quasi-linear scales for $f(R)$ gravity and nDGP, where the \emph{pseudo} nonlinear power spectrum is computed using either the \emph{pseudo} (solid lines) or `no-screening' (dashed lines) one-loop correction. Although the former approach is theoretically motivated, accuracy arguments make the latter preferable for all cosmologies and redshifts investigated in this work.

\section{Halo mass function and concentration tests}\label{sec:tests}
\begin{figure*}
\begin{center}
\includegraphics[width=\columnwidth]{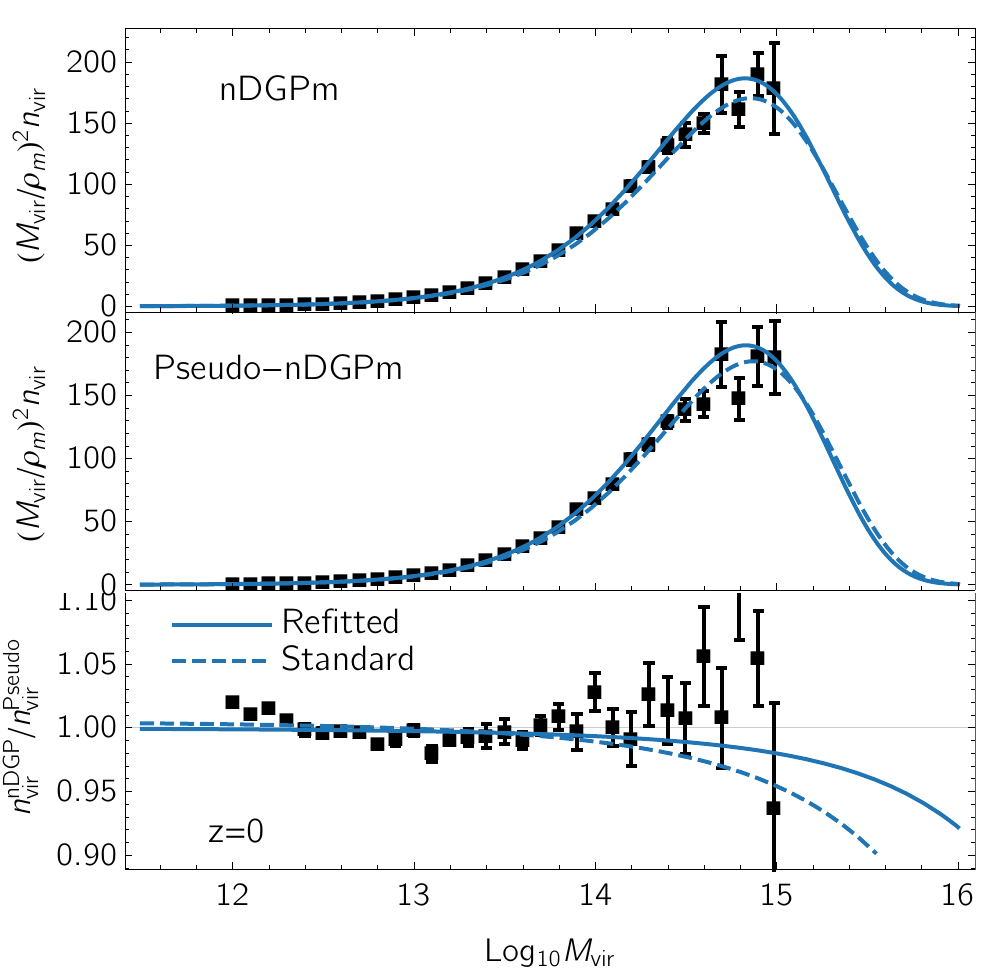}
\quad
\includegraphics[width=\columnwidth]{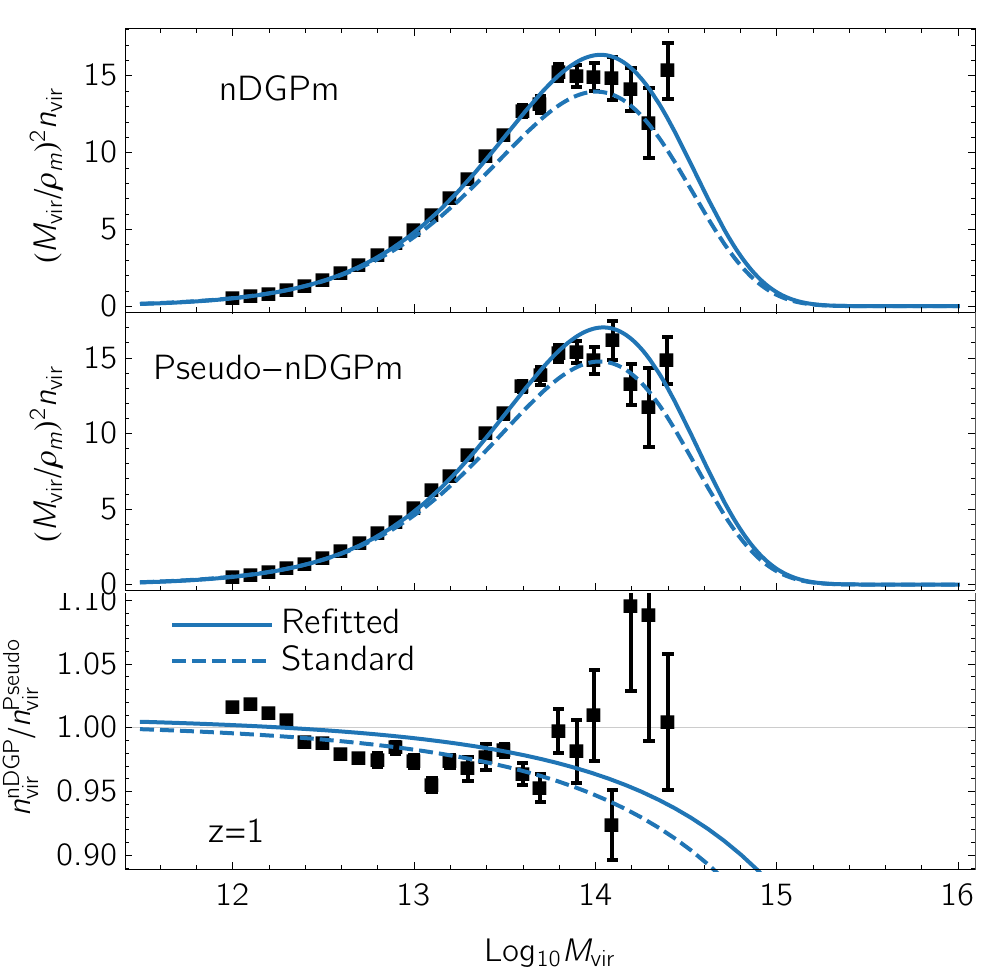}
\end{center}
\caption{$P_{1\rm h}(k \rightarrow 0)$ integrands (Eq.~\ref{eq:P1h}) at $z=0$ (left) and $z=1$ (right) for the nDGP cosmology with crossover scale $r_{\rm c} H_0=0.5$. The data points with error bars are Jackknife estimates from simulations; lines correspond to semi-analytical predictions using the standard Sheth-Tormen halo mass function fits $\{ A,q,p \}=\{0.3222,0.75,0.3\}$ (dashed), or to direct fits to our simulations (solid). Top panels show the results for the {\it real} nDGP cosmology, while middle panels present the outcome for the {\it pseudo} counterpart. The ratios of the {\it real} halo mass functions to the {\it pseudo} ones are illustrated in the lower panels. Combined with the information in Figure \ref{fig:respDGPm}, these ratios clearly stand out as the relevant quantity to achieve percent level accuracy for the reaction function over scales $k \lesssim 1$ \hMpc. In fact, the difference between the standard and refitted halo mass functions is more significant at low redshift, which reflects the performance shown in the lower panels of Figure \ref{fig:respDGPm}.}
\label{fig:P1hIgrDGPm}
\end{figure*}
\begin{figure*}
\begin{center}
\includegraphics[width=\columnwidth]{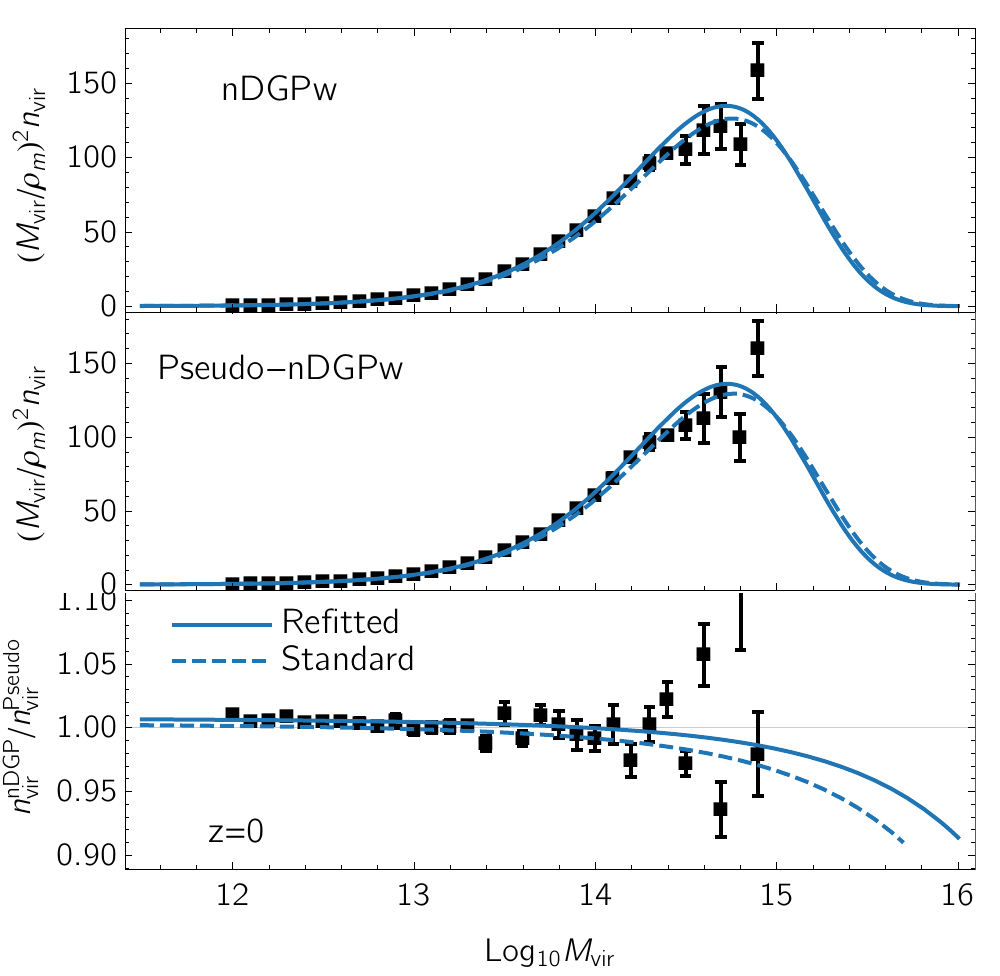}
\quad
\includegraphics[width=\columnwidth]{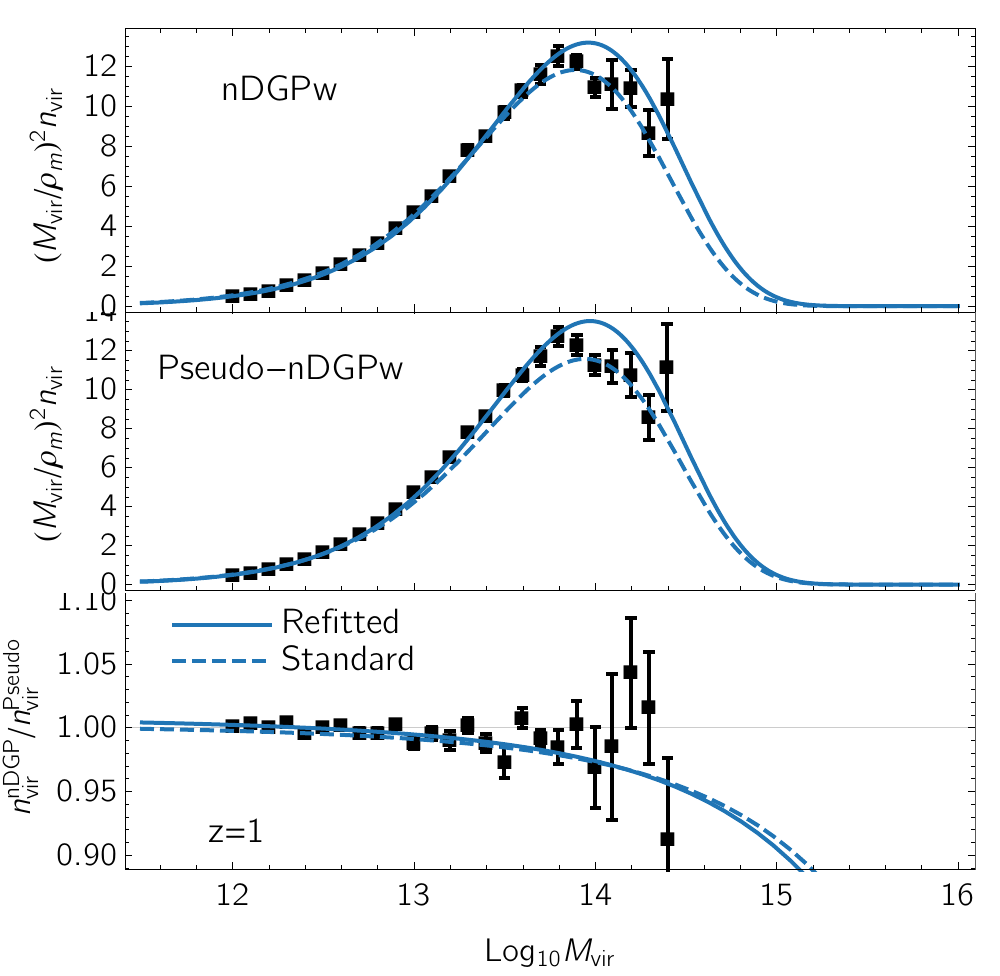}
\end{center}
\caption{Same as Figure \ref{fig:P1hIgrDGPm} for the nDGP cosmology with crossover scale $r_{\rm c} H_0=2$. For both selected redshifts the bottom panels show small differences between the standard and refitted halo mass function ratios. Once again, this matches the expectations from Figure \ref{fig:respDGPw}.}
\label{fig:P1hIgrDGPw}
\end{figure*}
\begin{figure}
\includegraphics[width=\columnwidth]{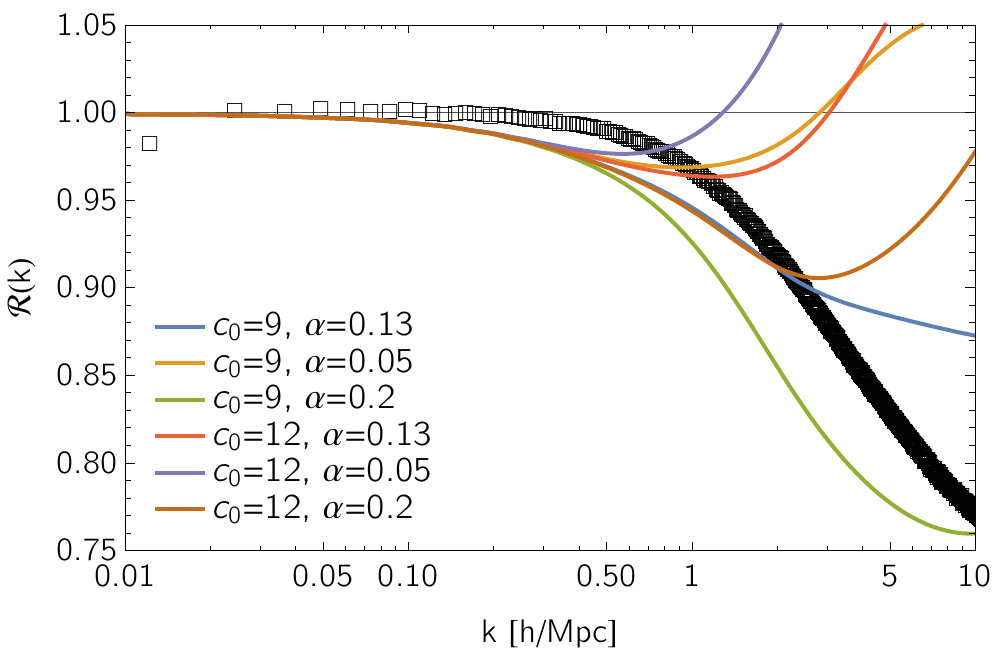}
\caption{Impact of the $c\text{-}M$ relation on the nDGPm halo model response at $z=0$. Lines correspond to different combinations of normalisation and slope in Eq.~\eqref{eq:concentration}, and changes are restricted to the \emph{real} cosmology only. Even significant deviations from the standard values (solid blue line) are not able to match the simulations, with virtually no effect on scales $k \lesssim 0.5$ \hMpc.}
\label{fig:respConc}
\end{figure}
\begin{table}
\begin{center}
\caption{Refitted Sheth-Tormen halo mass function parameters obtained from the simulation data shown in Figs.~\ref{fig:P1hIgrDGPm} and \ref{fig:P1hIgrDGPw}. Fits are consistent with $p=0$ in all cases. The standard values are $\{A, q, p\} = \{0.3222,0.75,0.3\}$.}
\begin{tabular}{c c c c}
\hline\hline
Model & $A$ & $q$ \\
\hline
nDGPm ($z=0$) 	& 0.3427 & 0.819  \\
nDGPm ($z=1$)	& 0.3067 & 0.757  \\
\hline
nDGPw ($z=0$) 	& 0.3347 & 0.819  \\
nDGPw ($z=1$)	& 0.3023 & 0.754  \\
\hline
pseudo-nDGPm ($z=0$) 	& 0.3438 & 0.829  \\
pseudo-nDGPm ($z=1$)	& 0.3057 & 0.761  \\
\hline
pseudo-nDGPw ($z=0$) 	& 0.3332 & 0.823  \\
pseudo-nDGPw ($z=1$)	& 0.3013 & 0.753  \\
\hline\hline
\end{tabular}
\label{tab:ST_refit}
\end{center}
\end{table}
In Sec.~\ref{sec:reactions} we showed that the \emph{real} to \emph{pseudo} halo mass function ratio, $n_{\rm vir}^{\rm real}/n_{\rm vir}^{\rm pseudo}$, controls the halo model reaction on scales $k \gtrsim 0.1$ \hMpc. In particular, for the nDGP cosmologies we explicitly checked that fitting the semi-analytical halo mass functions directly to our simulations results in better performance of the halo model reactions for wavenumbers $0.1 \lesssim k \, \Mpch \lesssim 1$. Below we briefly explain how these fits to the halo mass functions extracted from our simulations were carried out.

We measure the mean halo abundances and their uncertainties from simulations as outlined in~\cite{Cataneo:2016}, with the important difference that, here, halos identified with the {\sc rockstar} halo finder~\citep{Behroozi:2013} have masses defined in spherical volumes with mean density $\Delta_{\rm vir}$ times the background comoving matter density, as in Eq.~\eqref{eq:Mvir}. The virial overdensity depends on redshift, matter content and theory of gravity through Eq.~\eqref{eq:vir_overdensity}. Compared to~\cite{Cataneo:2016} we also use a finer mass binning, with a bin size $\Delta\log_{10}M = 0.1$. Figures~\ref{fig:P1hIgrDGPm} and~\ref{fig:P1hIgrDGPw} present a rescaled version of the halo mass functions (i.e. the large-scale limit of the one-halo integrand Eq.~\ref{eq:P1h}) as well as the $n_{\rm vir}^{\rm real}/n_{\rm vir}^{\rm pseudo}$ ratios. The standard halo mass function fits for the parameters in Eq.~\eqref{eq:SThmf} systematically underpredict the simulation measurements. Refitting these parameters to each \emph{real} and \emph{pseudo} simulation separately can change the halo mass function ratio $n_{\rm vir}^{\rm real}/n_{\rm vir}^{\rm pseudo}$ to an extent relevant for the halo model reactions. Note that when this ratio remains largely unaffected by the refitting, the predicted reactions also show very little variation (compare the right panels of Figure~\ref{fig:P1hIgrDGPw} and Figure~\ref{fig:respDGPw}). Table~\ref{tab:ST_refit} summarises the refitted ST halo mass function parameters for the \emph{real} and \emph{pseudo} nDGP cosmologies used in this work.

Given the structure of the one-halo term Eq.~\eqref{eq:P1h}, one might argue that inaccurate halo profiles are partially responsible for the few percent mismatch between the predicted and measured reactions in Figure~\ref{fig:respF5} and Figure~\ref{fig:respDGPm}. If that was the case, then having accurate halo mass functions would not necessarily correspond to having accurate halo model reactions. However, Figure~\ref{fig:respConc} shows otherwise, in that even extreme variations of the halo concentrations cause important changes only on scales $k \gtrsim 0.5$ \hMpc. 

From these tests we conclude that accurate virial halo mass function ratios are central to determining $\lesssim 1\%$ accuracy for the halo model reactions, and that the mean concentration of massive halos is well described by Eq.~\eqref{eq:concentration}, at least as long as this relation is employed in the reaction ratios. In future works we will investigate the implications  of refitting the $c$-$M$ relation to that measured in simulations, which has the potential to ameliorate our halo model predictions in the highly nonlinear regime.


\bsp	
\label{lastpage}
\end{document}